\documentclass[a4paper,11pt]{article}

\usepackage{amsmath,amssymb,bbm,graphicx,theorem}
\usepackage{epstopdf}
\usepackage{src/pdfsync}

\newcommand{\mvec}[1]{{\boldsymbol #1}}
\newcommand{\rvec}[1]{{\vec #1}}

\theoremstyle{break}
\theorembodyfont{\sffamily}

\newtheorem{Postulate}{Postulate}
\newtheorem{Theorem}{Theorem}

\newcommand{\jbox}[1]{
	~\\[0.3cm]
	~\hspace*{-0.0cm}\fbox{
		\begin{minipage}{0.96\textwidth}
			#1
		\end{minipage}
	}
	\\[0.3cm]
}

\begin{document}

\title{Basic Concepts for a Quantum Mechanical Theory of Events}

\author{Kim J. Bostr\"om}
\date{\today}

\maketitle

\begin{abstract}
A physical theory is proposed that obeys both the principles of special relativity and of quantum mechanics. As a key feature, the laws are formulated in terms of quantum events rather than of particle states. Temporal and spatial coordinates of a quantum event are treated on equal footing, namely as self-adjoint operators on a Hilbert space. The theory is not based upon Lagrangian or Hamiltonian mechanics, and breaks with the concept of a continuously flowing time. The physical object under consideration is a spinless particle exposed to an external potential. The theory also accounts for particle-antiparticle pair creation and annihilation, and is therefore not a single-particle theory in the usual sense. The Maxwell equations are derived as a straightforward consequence of certain fundamental commutation relations. In the non-relativistic limit and in the limit of vanishing time uncertainty, the Schr\"odinger equation of a spinless particle exposed to an external electromagnetic field is obtained.
\end{abstract}

\tableofcontents

\section{Prolog}

\subsection{Introduction}

It appears that all physical description is based upon the concept of time and space, since anything which is subject to our perception is located within time and space. The nature of these notions is so fundamental that in 1790 Immanuel Kant~\cite{Kant1790} conceived of time and space as the \emph{conditions of our perception} which are prior to any experience and which therefore cannot be influenced by objects which are only \emph{part} of our experience. He finds himself in agreement with Newton, who acclaimed in 1687 in his famous \emph{Principia}~\cite{Newton1790} the concept of absolute time and space as those entities that exist in themselves and without relation to anything external. 
This concept of time and space was firmly anchored in the minds of the physicists of the 19th century. It is therefore quite understandable why Einsteins Theory of Relativity~\cite{Einstein1905} has evoked so much resistance among physicists and philosophers. Einstein's theory broke with the concept of absolute time and space and demoted them into things which are influenced by the material world and by the state of motion of the observer; time and space lost their absolute status and became relative. 

The theory of Quantum Mechanics has then continued attacking the physical fundament by bringing to light that the position and the momentum of a particle, which formerly have been assumed to be independent quantities, are in fact intrinsically linked, so that there is the same amount of information in the particle's position as there is in its momentum. Even worse, these quantities are \emph{complementary}, so that the precise knowledge about one of them excludes the simultaneous precise knowledge about the other. 
And finally, the localization of a particle in space turned out to be in a principal way afflicted by an unavoidable finite uncertainty, which seemed to conflict with the classical concept of a particle as a mass point of zero extension. Hence, space ceased to be a classical entity and became \emph{quantum}.

Surprisingly though, time has survived the quantum mechanical revolution. In the non-relativistic quantum theory as well as in its relativistic counterpart, time is treated as a classical parameter and not as a quantized entity. There have been efforts to reformulate the quantum mechanical framework so that time would be put on the same footing as space, like the Theory of Relativity actually demands. But these approaches are controversal and do not seem to truly resolve the problem.
This amongst other inconsistencies has led many researchers to favour the ``peaceful coexistence'' of Relativity and Quantum Mechanics against a perfect integration of both theories into a single one. And indeed, the results of such peaceful coexistence are very convincing. The Quantum Field Theory leads to incredibly precise predictions and satisfies both quantum mechanical and special-relativistic demands, although there still appear certain annoying infinities that have to be removed by hand from the calculations. Under certain circumstances, and in particular when trying to incorporate the gravitational aspect of Relativity, the infinities can no longer be removed, and it remains unclear how a unified theory should tackle with this problem.
\medskip

In the present paper, which is based on previous work~\cite{Bostroem03c}, a quantum mechanical theory is put forward which incorporates in its fundament the concepts of Quantum Mechanics and of Special Relativistic Mechanics. At the present stage, the theory requires a flat spacetime structure. It should, however, be possible in principle to find a formulation for curved spacetime. 

The proposed theory can also be considered as a novel approach to the problem of a time operator in quantum mechanics, and it shows the following features:
\begin{enumerate}
\item
It entails a quantization of time in full analogy to space.
\item
It is relativistic. 
\item
It is not based upon Lagrangian or Hamiltonian mechanics.
\item
It implies the standard theory as a limit case. 
\item
It is not afflicted by divergencies. 
\end{enumerate}
The physical object that is described by the theory is a spinless particle in empty space, exposed to an external potential. Yet, the theory also accounts for the creation and annihilation of a particle-antiparticle pair, and is therefore not a single-particle theory in the usual sense. 

The solution to this apparent contradiction lies in a new concept of spacetime kinematics. Instead of putting the state of a system, here an individual particle, into the focus of the theory, it is the individual \emph{event} which is the fundamental entity. The theory confines itself to the way that the particle comes into being for the perception of a potential observer, namely in form of an event with time and space among its observable qualities. 
Since the event is the most fundamental object in the presented quantum mechanical theory, it will be called the Quantum Event Theory (QET).

The theory does not contain any background time or proper time that would still represent a classical dynamical parameter. Rather, time appears merely as one particular component of the self-adjoint 4-position operator attributed to a single event. Hence, temporal and spatial coordinates are all described in operator form, and the coordinate axes represent the continuous spectra of these operators. 
Since time is described by a quantum mechanical observable with a purely continuous spectrum it shows an unavoidable finite uncertainty, and therefore it becomes impossible to further conceive of a \emph{point} in time where something actually is existing. Clearly, this conflicts with the assumption of a continuous flow of time. 

The theory is not based upon Lagrangian or, equivalently, Hamiltonian mechanics, because these are based on the principle of least action and require the existence of state trajectories and therefore rely on the concept of a flowing time, which is not possible when considering time as a quantum number. Hamiltonian mechanics appears as a limit case of QET when one considers a vanishing time uncertainty under non-relativistic conditions.    

We will work in a unit system where no natural constants appear in the equations and where the only unit is the unit of distance. Later on, we will change to the SI unit system by implementing the historically motivated natural constants.

We will derive the Maxwell equations as a straightforward consequence of the fundamental commutation relations between the components of the 4-position and 4-momentum operators. In light of this derivation, the electromagnetic interaction appears as a profound expression of the quantum nature of time and space.

By applying the nonrelativistic limit and the limit of vanishing time uncertainty, we will eventually derive the Schr\"odinger equation of a spinless particle exposed to an electromagnetic field.

\subsection{Provocative thoughts: A new role for space and time?}

Let us have a look at the famous part of Newton's \emph{Principia}~\cite{Newton1790} where he introduces the notion of absolute time:

\begin{quote}\em
Absolute, true, and mathematical time, of itself and from its own nature, flows equably and without relation to anything external.
\end{quote}

Now let us slightly modify the above formulation:

\begin{quote}\em
True and mathematical time, of itself and from its own nature, flows equably.
\end{quote}

We obtain a statement which would be true within all generally accepted physical theories. Let us denote the so-defined concept of physical time the \emph{flowing time}. In particular, the concept of flowing time implies that there are no ``holes'' within the evolution of a system. In other words:

\begin{quote}\em
A physical system is at any point in time in a particular state.
\end{quote}

Now this is indeed the basis of our intuition and of all generally accepted physical theories. The change of state of a physical system in the course of the flowing time represents the \emph{continuous evolution} of the system. A physical theory should then be capable of telling us how this evolution takes place in that it predicts or retrodicts the state of the system at a certain time point conditional on the state of the system at another time point. This condition then represents the final or initial condition, depending on the ordering between the two time points.

QET breaks with this fundamental concept of flowing time. The continuous evolution of a system is replaced by the discrete appearance of individual events. 
Between neighbouring events there ``is'' nothing, because nothing is happening. The notion of \emph{existence} which seems to be already at the fundament, appears as a manifestation of a more fundamental notion: \emph{occurrence}. When an electron does not show up in form of a detector event, its actual existence is a mere assumption. On the other hand, whenever a detector event occurs, the electron is certain to exist at the time of occurrence. Hence, occurence gives rise to existence, and existence is in need of occurence. From this point of view it appears reasonable to formulate a physical theory on the basis of the factual occurence of individual events instead of the assumed existence of a definite system state. 

Two neighbouring events constitute an indivual quantum process.
QET enables us to calculate the probability for the occurence of any such process. Since it is a purely probabilistic theory, it does not have to face the interpretational ambiguities usually addressed as the ``measurement problem'', which basically stem from the coexistence of a deterministic and continuous evolution, and a probabilistic and discontinuous state reduction. QET decides in favour of the probabilistic and discontinuous aspects. The factual history of the universe is entirely composed from discrete quantum events that cause one another in a probabilistic manner. Although the underlying laws of nature are clear and unambiguous, they are purely probabilistic, hence the unravelling of an individual history is not uniquely determined by the first event. 

\begin{figure}[t]
	\[\includegraphics[width=0.7\textwidth]{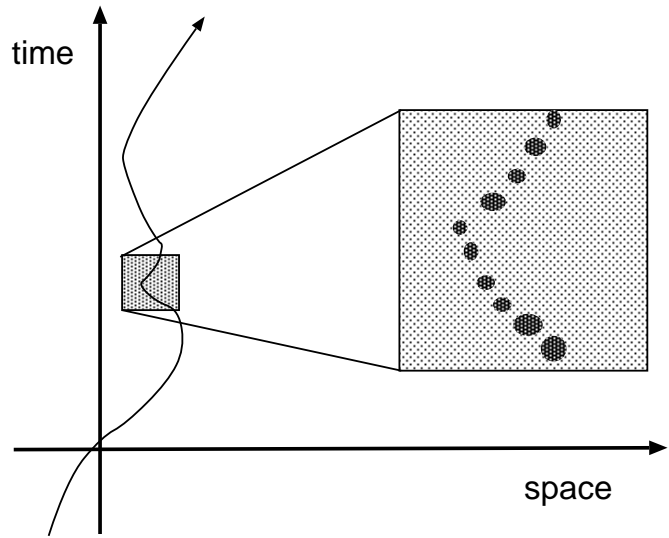}\]
		\vspace*{-0.5cm}\caption{\small A continuous worldline of a classical particle. In QET every worldline decomposes into discrete elementary events of finite size. The theory then gives the probability for the transition between neighbouring events. The size of each event depends on the resolution of the corresponding detector. For example, the gas molecules in a fog chamber may serve as detectors.}\label{worldline}
\end{figure}

\begin{figure}[t]
	\[\includegraphics[width=0.7\textwidth]{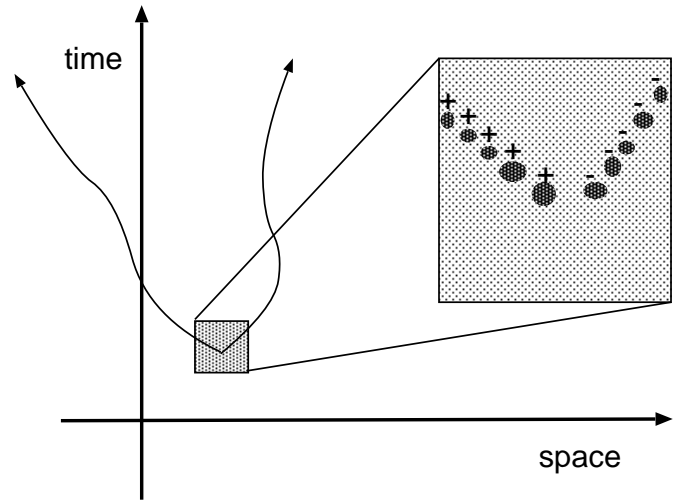}\]
		\vspace*{-0.5cm}\caption{\small The creation of a particle-antiparticle pair. The worldline of each particle abruptly ends at the point of creation. In QET there would be no such abrupt ending. Instead, the creation process is described as a series of jumps of a single event in spacetime, switching the sign of energy and thereby the direction of time.}\label{creation}
\end{figure}

Clearly, a single event does not evolve, it simply \emph{happens}. Therefore, the time operator appearing in the theory does not measure a \emph{dynamical} quantity, in that it does not quantify \emph{change} but rather a component of \emph{position in spacetime}. Dynamics enters the theory only in form of discrete quantum processes each involving two quantum events. If one event occurs in the lightcone of the other then the process appears to the observer as a change of state of one particle (see Fig.~\ref{worldline}). If both events occur outside each other's lightcone then the process looks like the detection of two particles with different properties. This appears to the observer as the creation or annihilation of a pair of particles (see Fig.~\ref{creation}). 
QET in its present form is not a single-particle but rather a \emph{single-event} theory, because it describes the history of one individual event on its journey through space and time. The event history, however, is not necessarily coupled to the direction of time, which distinguishes it from the particle history as it appears to the observer. There is a kind of ``meta-ordering'' between the events, that is not linked to the course of time but rather to the course of the stochastic process. 

Time is not considered as a \emph{dynamical} entity but rather as one coordinate of an event, and treated in the same manner as the spatial coordinates. Coordinates of an event do not undergo any sort of evolution, they do not change, they simply \emph{are}, because they only describe one single event which \emph{occurs} somewhere in space and time. 

On the other hand, the time operators which are usually considered in the literature account for the \emph{change} of the system, and are thus in some way or the other dependent on the external conditions. Let us call operators of this kind \emph{dynamical time operators}. It is not necessarily so that the dynamical time operators are all explicitly time-dependent. In fact, many dynamical time operators are not time-dependent, for example the arrival time operators. But all dynamical time operators are intertwined with the external setting. If the external setting is altered, e.g. some energetic potential is switched on or off, the dynamics of the system will also alter, and so will the dynamical time operator. Not so with the time operator considered in QET. It is not dependent in any way on the external setting, on the existence  of forces, potentials, perturbations, interactions or whatever. It remains completely unaffected, because it represents \emph{the} time.

What is \emph{the} time? What is \emph{the} space? In Kant's view time and space are the conditions of perception, they are the most fundamental entities which all of our perception and therefore all of our description is build upon.
In QET the most fundamental entity is not time and space, it is the event. And time and space are only specific \emph{qualities} of an event, just like color and taste are specific qualities of an apple. 
Kant is, and most of us are, visualizing time and space as the big arena where everything happens. Let us reverse this picture: The things that happen, the events itself, are the arena for time and space! Each event \emph{creates} its own time and space, and through the occurrence of all these events, the spacetime is unfolded and blown up like a balloon that contains the evolving universe.

Possibly, time has not yet been \emph{truly} quantized because it represents the last fibre that ties quantum mechanics to the classical intuition of a continuous spacetime arena. 
I believe that as long as we do not cut this fibre, we will not succeed in finding a grand unified theory. 

\begin{figure}[t]
	\[\includegraphics[width=0.6\textwidth]{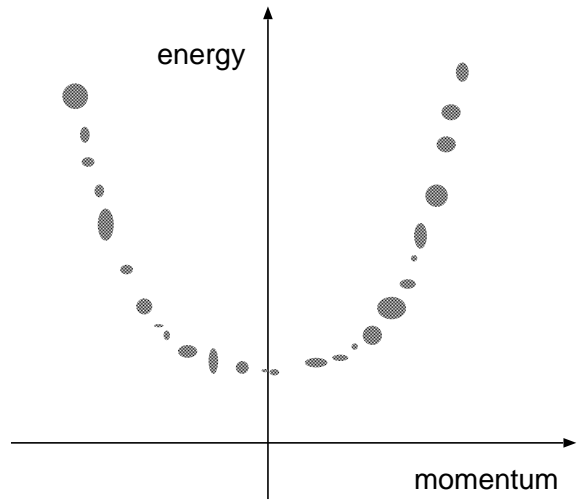}\]
		\vspace*{-0.5cm}\caption{\small The history of an event as it would look in the kinetic energy-momentum representation. A fundamental principle of QET demands that all events that constitute the history are located on the kinetic mass shell (here only the upper part is shown). In the total energy-momentum representation, the history would in general look more complicated, because the mass shell is deformed by the potentials.}\label{pHistory}
\end{figure}

In QET we consider the spacetime coordinates as just certain qualities of an event, and the coordinate axes are just the continuous spectra of the corresponding time and space operators. If we take other operators then we could describe everything that happens without ever mentioning time or space. We could, for example, tell the history of the universe completely in terms of energy and momentum (see Fig.~\ref{pHistory}). 

One might say that a quantum state cannot be as fundamental as a quantum event because it does not entail intrinsic information about time. Within the framework of any theory whose central object of description is the state of a system, time is always an external degree of freedom which stems from the classical external environment (us, the observer). The question is: Does Nature truly consider time an external classical parameter? This might be the case, and the existing theories tell us so. But maybe this is not the case, and then we need a new theory.

\subsection{Some theoretical context}

There is a long history of controversies about the role of time in quantum mechanics, and a glimpse at the amount of controversal literature on this issue makes clear that proposing a ``new'' treatment of time is much like taking a fresh jump into the mud. I will not comment on all the currently existing approaches, not only because it would take too much room, but also because there already are much better reviews than I could possibly provide. For example, a very good review on the role of time in quantum mechanics can be found in \cite{Muga02}. 

In the previous section it should have become clear that I am not considering dynamical time operators as, for example, the operators for the time of arrival~\cite{Delgado97, Muga98, Egusquiza99, Werner87, Oppenheim99}, the time of flight~\cite{Moshinsky52, Szriftgiser96}, the time of occurence~\cite{Brunetti02a}, the lifetime~\cite{Smith60}, the tunneling time~\cite{Hauge89, Landauer94}, and the characteristic time~\cite{Mandelstamm45, Muga02a}. I am also not considering the duration and timing of quantum jumps~\cite{Weisskopf30, Dalibard92, Plenio98}, nor am I involved in quantum clocks and their behaviour~\cite{Mandelstamm45, Hartle88, Salecker58, Buzek99}. The conception probably closest to mine seems to be the \emph{consistent histories approach} \cite{Griffiths84, Omnes92, Hartle88, Hartle91}, which describes the evolution of a quantum system as a series of individual events. However, in this approach the time is a sharply defined classical parameter and the dynamics is unitary. Another approach similiar to QET seems to be at first sight the \emph{event enhanced quantum theory} (EEQT)~\cite{Blanchard96, Blanchard00}. However, a closer look shows that the events considered in EEQT are associated with the changes of the state of a classical system coupled to a quantum system, and are modeled on the basis of a modified Schr\"odinger equation involving a classical time parameter. Also the relativistic version of EEQT is based on a classical time parameter, namely the proper time. 

The above mentioned approaches are based on some sort of classical time parameter and involve piecewise continuous evolution. This goes in contrast to the theory proposed here, where time is considered as merely one particular coordinate of an individual event, and is quantized in exactly the same manner as the spatial coordinates. The dynamics predicted by the theory is not continuous, and it is not unitary, except in the non-relativistic limit where particle creation and annihilation is excluded.

\section{Axiomatic Foundation}

\subsection{Quantum Events}

Following the reasoning of the previous section, instead of assuming that the universe is composed out of elementary particles, we assume that the history of the universe is composed out of elementary events which we will also call \emph{quantum events}. In this way, not only the spatial but also the temporal aspect of the universe obtains a discrete, grainy structure. There is no continuous evolution like in the standard theories, but rather a discrete sequence of individual transitions, or ``jumps'', from one quantum event to another. The theory which we will be developing here enables us to calculate the probability for each individual jump process.

Classically, a particle appears at some time $t$ and at some position $\rvec x$, so the corresponding quantum event would be represented by a point $(t,\rvec x)$ in Minkowski spacetime $\mathbbm R^4$ which is called the 4-position. Let us use the shorthand notation
\begin{equation}
	\mvec x \equiv (t,\rvec x)
\end{equation}
for points in spacetime, and fix the natural unit of each component of ${\mvec x}$ to ``1 meter', in short ``1 ${\rm m}$'',
\begin{equation}
	[x^\alpha]=1\,{\rm m}.
\end{equation}
The zeroth component $x^0=t$ obtains the same unit as the spacelike components $\rvec x$, which is understood in the sense that $t=10\,{\rm m}$ denotes the time that light needs to travel a distance of 10 meters through the empty space. (One might call this unit of time a ``lightmeter'', in analogy to the the ``lightyear'' as a unit of distance.) By construction then, the speed of light is $c=1$.

Now it is straightforward to represent the quantum version of a quantum event by a complex wave packet $\psi(\mvec x)$ in Minkowski spacetime.
However, in contrast to the wave function of standard quantum mechanics, let us understand the zeroth component $x^0=t$ of the argument of $\psi$ not as the parameter of continuous time evolution, but rather as the temporal coordinate of a single quantum event, just like $\rvec x$ are the spatial coordinates. 
Let us call $\psi(\mvec x)$ the \emph{event wave function}. In direct analogy to the standard quantum theory, we require that the integration of $|\psi(\mvec x)|^2$ over spacetime yields a finite value,
\begin{equation}\label{Psinorm}
	\int d^4x\,|\psi(\mvec x)|^2<\infty.
\end{equation}
In other words, any event wave function $\psi(\mvec x)$ is an element of the Hilbert space ${\cal E}:=L^2(\mathbbm R^4)$ of square-integrable functions over spacetime, which will be called the \emph{event space}, and which is endowed with the scalar product
\begin{equation}\label{scalarprod}
	\langle\psi|\phi\rangle:=\int d^4x\,\psi^*(\mvec x)\phi(\mvec x),
\end{equation}
so that the norm in ${\cal E}$ is given by $\|\psi\|=\sqrt{\langle\psi|\psi\rangle}$.
All this can be collected into the following postulate.

\jbox{
\begin{Postulate}[Quantum Events]\label{Events}
The history of the universe is composed out of quantum events.
Each quantum event indicates the detection of a particle. A spinless quantum event is mathematically represented by an element $\psi$ of the Hilbert space ${\cal E}=L^2(\mathbbm R^4)$.
\end{Postulate}
}
Let us switch to the abstract Dirac notation. 
In a process of elementary quantization, the spacetime coordinates of a classical event $\mvec x$ are mapped to a ket $|\mvec x\rangle$,
\begin{equation}\label{quant}
	\mvec x\mapsto |\mvec x\rangle,
\end{equation}
so the ket $|\mvec x\rangle$ can be regarded as a quantum version of the classical event $\mvec x$. As it stands for a point event, the ket $|\mvec x\rangle$ is an idealized entity.
A realistic quantum event $|\psi\rangle$ would be spread over a finite region of uncertainty, so it is represented by a continuous linear superposition of point events
\begin{equation}
	|\psi\rangle=\int d^4x\,\psi(\mvec x)|\mvec x\rangle,
\end{equation}
where the complex function $\psi(\mvec x)$ is an element of the Hilbert space $L^2(\mathbbm R^4)$. In this abstract picture, the event space populated by the vectors $|\psi\rangle$ is the ket space
\begin{equation}
	{\cal E}=\big\{\int d^4x\,\psi(\mvec x)|\mvec x\rangle\mid
		\psi\in L^2(\mathbbm R^4)\big\}.
\end{equation} 
We shall refer to the set
\begin{equation}
	{\cal B}_x:=\{|\mvec x\rangle\mid \mvec x\in\mathbbm R^4\}
\end{equation}
as the \emph{4-position basis} whose members fulfill the orthonormal and completeness relations
\begin{eqnarray}
	\langle\mvec x|\mvec x'\rangle&=&\delta(\mvec x-\mvec x')\\
	\int d^4x\,|\mvec x\rangle\langle\mvec x|&=&\mathbbm 1.\label{xcomplete}
\end{eqnarray}
The structure of the Hilbert space ${\cal E}$ guarantees that any non-trivial vector $|\psi\rangle\in{\cal E}$ is extended over time and space, because the completeness of the Hilbert space implies that any vector extended over a set of measure zero coincides with the null vector denoted by $|\emptyset\rangle$. The finite spacetime extension of any nontrivial vector in ${\cal E}$ implies an uncertainty in time and space.
Although the point events $|\mvec x\rangle$ are not extended, they are, as basis vectors, assumed to be \emph{not} identical to the null vector. 
This is only possible because the point events $|\mvec x\rangle$ are in fact \emph{improper} vectors outside the Hilbert space ${\cal E}$.
They belong to the \emph{distribution space} $\Theta^\dag$ which is the dual to the \emph{test space} $\Theta\simeq{\cal S}(\mathbbm R^4)$, where ${\cal S}(\mathbbm R^4)$ is the \emph{Schwartz space} of rapidly decreasing functions over $\mathbbm R^4$.
While $\Theta,{\cal E}$ and $\Theta^\dag$ are \emph{abstract} spaces, the corresponding spaces ${\cal S}(\mathbbm R^4),L^2(\mathbbm R^4)$ are \emph{function} spaces and the space ${\cal S}^\dag(\mathbbm R^4)$ is the space of linear-continuous \emph{functionals} over ${\cal S}(\mathbbm R^4)$.
The test space $\Theta$, Hilbert space ${\cal E}$ and distribution space $\Theta^\dag$ form a \emph{Gelfand triplet}~\cite{Bohm89}: 
\begin{equation}
	\Theta\subset{\cal E}\subset\Theta^\dag.
\end{equation}
With $|\chi\rangle\in\Theta^\dag$ being an improper vector, the bracket $\langle\chi|\vartheta\rangle$ is for any test vector $|\vartheta\rangle\in\Theta$ defined as the complex number
\begin{equation}
	\langle\chi|\vartheta\rangle:=\chi^*[\vartheta],
\end{equation}
where $\chi:{\cal S}(\mathbbm R^4)\rightarrow\mathbbm C$ is the distribution corresponding to $|\chi\rangle$ and $\vartheta$ is the test function corresponding to $|\vartheta\rangle$.

\subsection{Properties}

Classical properties can be represented by sets. An object $x$ is said to have a property $a$ if there is a corresponding set ${\cal M}_a$ so that $x\in{\cal M}_a$. The \emph{type} of the object is represented by a set $\Omega$ so that any allowed property of type-$\Omega$ objects is represented by a subset of $\Omega$. The logical operations are represented by set operations, namely
\begin{equation}\label{clogic}
\begin{split}
	a\wedge b &\quad\leftrightarrow\quad {\cal M}_a\cap{\cal M}_b\\
	a\vee b &\quad\leftrightarrow\quad {\cal M}_a\cup{\cal M}_b\\
	\sim a &\quad\leftrightarrow\quad {\cal M}_a^c
		\equiv\Omega\setminus{\cal M}_a,
\end{split}
\end{equation}
where $a\wedge b$ means ``$a$ and $b$'', $a\vee b$ means ``$a$ or $b$'' and $\sim a$ means ``not $a$''.

Quantum properties can be represented by vector spaces, so that an object $\psi$ has the property $a$ if there is a corresponding vector space ${\cal V}_a$ so that $\psi\in{\cal V}_a$. The type of the object is represented by a Hilbert space $\Omega$ so that any allowed property of type-$\Omega$ objects is represented by a subspace of $\Omega$. 
The logical operations are represented by vector space operations, namely
\begin{equation}\label{qlogic}
\begin{split}
	a\wedge b &\quad\leftrightarrow\quad {\cal V}_a\cap{\cal V}_b\\
	a\vee b &\quad\leftrightarrow\quad {\cal V}_a+{\cal V}_b\\
	\sim a &\quad\leftrightarrow\quad {\cal V}_a^\bot,
\end{split}
\end{equation}
where the sum of two subspaces ${\cal V}_a,{\cal V}_b\subset\Omega$ is defined by
\begin{equation}
	{\cal V}_a+{\cal V}_b:=\{\psi+\phi\mid\psi\in{\cal V}_a,
		\phi\in{\cal V}_b\},
\end{equation}
and where the orthogonal complement of a subspace ${\cal V}_a\in\Omega$ is defined by
\begin{equation}
	{\cal V}_a^\bot:=\{\phi\in\Omega\mid\langle\phi|\psi\rangle=0,
		\forall\psi\in{\cal V}_a\}.
\end{equation}
The algebraic structure~\eqref{qlogic} is also referred to as ``quantum logic''.  

The objects of standard quantum mechanics are elements of a Hilbert space ${\cal H}=L^2(\mathbbm R^3)$. The type of these objects is that of the state of a spinless particle. The objects of QET are elements of the Hilbert space ${\cal E}=L^2(\mathbbm R^4)$ and so they are of a different type. Indeed, they do not represent states of a particle but rather individual detector events. The bridge between events and particles is given by postulate~\ref{Events}: Each event indicates the detection of a particle. The properties of particles are therefore inherited by the properties of events, which are defined by the subspaces of ${\cal E}$ as constituted by the following postulate.
\jbox{
\begin{Postulate}[Properties and Detectors]\label{Properties}
Properties of quantum events are represented by subspaces of the event space ${\cal E}$.
A quantum event $\psi\in{\cal E}$ is said to show the property $a$ if there is a corresponding subspace ${\cal V}_a\subset{\cal E}$ so that $\psi\in{\cal V}_a$.
The logical operations are realized by~\eqref{qlogic}.

A \emph{detector} for the property value $a$ corresponding to the subspace ${\cal V}_a$ is represented by the projector $\hat\Pi_a$ projecting on ${\cal V}_a$. 
\end{Postulate}
}
As follows from the logical relations~\eqref{qlogic},
two properties $a$ and $b$ are mutually exclusive exactly if their corresponding detectors obey $\hat\Pi_a\hat\Pi_b=0$. 
An event $\psi$ has the property $a$ exactly if $\hat\Pi_a|\psi\rangle=|\psi\rangle$. If ${\cal A}=\{a\}$ is a discrete set of real numbers, and $\{\hat\Pi_a\}$ is a corresponding set of mutually orthogonal projectors, then each value $a$ can be interpreted as a possible value of some observable aspect represented by the self-adjoint operator
\begin{equation}
	\hat A=\sum_{a\in{\cal A}} a\,\hat\Pi_a,
\end{equation}
so that $\psi$ has the property $a$ exactly if it is an eigenvector of $\hat A$ with the eigenvalue $a$, i.e. $\hat A|\psi\rangle=a|\psi\rangle$.
The set ${\cal A}$ is then identical to the discrete spectrum $\sigma(\hat A)$.
Let us now consider \emph{generalized projectors} which project onto \emph{improper
subspaces} of ${\cal E}$. 
If $\hat A$ is a self-adjoint operator with a continuous spectrum $\sigma(\hat A)$, and if $a\in\sigma(\hat A)$, then
\begin{equation}
	\hat\Pi_a=\delta(\hat A-a)
\end{equation} 
is the generalized projector onto the improper eigenspace ${\cal V}_a$ corresponding to the eigenvalue $a$. The generalized projectors are mutually orthogonal,
\begin{equation}
	\hat\Pi_a\hat\Pi_{a'}=\delta(a-a')\hat\Pi_a,
\end{equation}
and complete
\begin{equation}
	\int_{\sigma(\hat A)}da\,\hat\Pi_a=\mathbbm1,
\end{equation}
so that there is an essentially self-adjoint operator $\hat A$ defined by
\begin{equation}\label{Acont}
	\hat A=\int_{\sigma(\hat A)}da\,a\,\hat\Pi_a.
\end{equation}
The improper eigenspace ${\cal V}_a$ is in fact \emph{not} a subspace of ${\cal E}$, but rather it is a subspace of the distribution space $\Theta^\dag$ from the Gelfand triplet $(\Theta,{\cal E},\Theta^\dag)$. The correct definition of ${\cal V}_a$ is therefore
\begin{equation}
	{\cal V}_a=\{\chi_a\in\Theta^\dag\mid (\hat A-a)|\chi_a\rangle=0\},
\end{equation}
so any element of ${\cal V}_a$ is a distribution on the test space $\Theta\subset{\cal E}$.
Because ${\cal V}_a$ is not a subspace of ${\cal E}$, it does, by Postulate~\ref{Properties}, not represent a property of type-${\cal E}$ objects, that is, of quantum events. So what is the intepretation in this case?
Although any $\hat\Pi_a$ is an improper projector, any \emph{integral} over a subset ${\cal A}\subset\sigma(\hat A)$ yields a proper projector,
\begin{equation}
	\hat\Pi_{\cal A}:=\int_{\cal A}da\,\hat\Pi_a.
\end{equation}
Therefore, the space ${\cal V}_{\cal A}$ corresponding to the projector $\hat\Pi_{\cal A}$ \emph{is} a subspace of ${\cal E}$ and thus ${\cal A}$ represents a valid property of quantum events. One should note that the operator $\hat A$ defined by~\eqref{Acont} is only \emph{essentially} self-adjoint, which means that it is symmetric on a dense subset of ${\cal E}$, and has a unique self-adjoint extension. We combine the above reasonings into the following postulate.
\jbox{
\begin{Postulate}[Observables]\label{Observables}
Any essentially self-adjoint operator $\hat A$ on ${\cal E}$ is called an \emph{observable} and represents an observable quality of a quantum event with the spectrum $\sigma(\hat A)$ as the set of observable values. The value
\begin{equation}
	\langle\hat A\rangle:=\frac{\langle\psi|\hat A|\psi\rangle}
		{\langle\psi|\psi\rangle}
\end{equation}
is called the \emph{center} of $\hat A$ for the event $\psi\in{\cal E}$, and the value
\begin{equation}
	\Delta A:=\sqrt{\langle\hat A^2\rangle-\langle\hat A\rangle^2}
\end{equation}
is called the \emph{uncertainty} of $\hat A$ for the event $\psi$.
\end{Postulate}
}
The values $\langle\hat A\rangle$ and $\Delta A$ are \emph{characterizations} of $\psi$ with respect to $\hat A$ and are not understood in a statistical sense. Statistics will enter the theory in form of transition probabilities between individual events.
Together with Postulate~\ref{Properties} we see that if the uncertainty vanishes, $\Delta A=0$, then $\psi$ is an eigenvector of $\hat A$, so that the corresponding eigenvalue $a$ coincides with the center $\langle\hat A\rangle$, so that $\langle\hat A\rangle=a$ can be interpreted as the exact measured value of the property represented by the eigenspace ${\cal V}_a$ corresponding to $a$.
If $\hat A$ is an observable with a purely continuous spectrum, then there is no eigenvector in ${\cal E}$ and we have $\Delta A>0$ for any $\psi\in{\cal E}$, so there is always a finite uncertainty of $\hat A$. Only the subspaces containing $\psi$ represent the actual properties of $\psi$. If $\hat\Pi_a$ is the projector onto the subspace ${\cal V}_a$, then $\psi$ shows the corresponding property $a$ exactly if
\begin{equation}
	\langle\hat\Pi_a\rangle=1,
\end{equation}
and therefore
\begin{eqnarray}
	\Delta\Pi_a
		&=&\langle\hat\Pi_a^2\rangle-\langle\hat\Pi_a\rangle^2\\
		&=&\langle\hat\Pi_a\rangle-\langle\hat\Pi_a\rangle^2=0.
\end{eqnarray}
Let $\lambda$ be any real number, $\hat A,\hat B$ self-adjoint operators on ${\cal E}$ and $\psi\in{\cal E}$ a normalized vector. Then the operators $\hat A'\equiv\hat A-\langle\hat A\rangle$ and $\hat B'\equiv\hat B-\langle\hat B\rangle$ are also self-adjoint and thus
\begin{eqnarray}
	0&\leq&\langle\psi|(\hat A'-i\lambda\hat B')
		(\hat A'+i\lambda\hat B')|\psi\rangle\\
	&=&\langle\psi|(\hat A'^2+i\lambda[\hat A',\hat B']
	+\lambda^2\hat B'^2)|\psi\rangle\\
	&=&\lambda^2\langle\hat B'^2\rangle+\lambda\langle i[\hat A,\hat B]\rangle
	+\langle\hat A'^2\rangle\\
	&=&\lambda^2(\Delta B)^2+\lambda\langle i[\hat A,\hat B]\rangle+(\Delta A)^2.
\end{eqnarray}
The righthand side of the above inequality is a quadratic function in $\lambda$ of the form $f(\lambda)=a\lambda^2+b\lambda+c$, which is only nonnegative if the coefficients fulfill $b^2\leq 4ac$, thus
\begin{eqnarray}
	\langle i[\hat A,\hat B]\rangle^2&\leq&4(\Delta B)^2(\Delta A)^2
\end{eqnarray}
from where one obtains the \emph{Heisenberg uncertainty relation}
\begin{equation}
	\Delta A\,\Delta B\geq \frac12|\langle[\hat A,\hat B]\rangle|.
\end{equation}
Consequently, an event cannot show two properties $a$ and $b$ if the projectors onto the corresponding subspaces do not commute. Note that additions like ``at the same time'' or ``simultaneously'' are not necessary here, yet they are even meaningless, because time is not involved as a special parameter. An event shows a property or not, and that is all.

\subsection{Position and momentum}
\nopagebreak
\jbox{
\begin{Postulate}[Position and Momentum]\label{Position}
The \emph{4-position} of a quantum event is represented by a 4-dimensional operator $\hat{\mvec x}$ defined by
\begin{equation}
	\hat{\mvec x}\psi(\mvec x)=\mvec x\psi(\mvec x).
\end{equation}
The dynamics is governed by a 4-dimensional observable $\hat{\mvec\wp}$ called the \emph{kinetic 4-momentum}, which is a self-adjoint operator on ${\cal E}$ whose components obey the commutation relations
\begin{eqnarray}\label{kmomentum}
	[\hat \wp^\alpha,\hat x^\beta]&=&i g^{\alpha\beta}.\label{Px0}
\end{eqnarray}
The commutators of the components of $\hat{\mvec\wp}$ define the components of the self-adjoint tensor operator
\begin{equation}
	\hat F^{\alpha\beta}:=i[\hat \wp^\alpha,\hat \wp^\beta],\label{PP0}
\end{equation}
which is called the \emph{interaction tensor}. The case $\hat F^{\alpha\beta}=0$ is denoted as the \emph{free case}.
The \emph{total 4-momentum} $\hat{\mvec p}$ is defined by
\begin{equation}
	\hat{\mvec p}\psi(\mvec x):=i\mvec\partial\psi(\mvec x),\label{p}
\end{equation}
and the \emph{potential 4-momentum} is defined by
\begin{equation}
	\hat{\mvec A}:=\hat{\mvec p}-\hat{\mvec\wp}.\label{A}
\end{equation}
\end{Postulate}
}
In analogy to the Hamilton operator $\hat H$ in standard QM, the kinetic 4-momentum $\hat{\mvec\wp}$ must be apriori known to make any predictions about the dynamics. 
Just like with the kinetic 4-momentum $\hat{\mvec\wp}$, the components of the total 4-momentum $\hat{\mvec p}$ fulfill the relations
\begin{equation}
	[\hat p^\alpha,\hat x^\beta]=i g^{\alpha\beta},\label{xp0}
\end{equation}
but in contrast to $\hat{\mvec\wp}$ they all mutually commute,
\begin{equation}
	[\hat p^\alpha,\hat p^\beta]=0.\label{pp0}
\end{equation}
Therefore the total 4-momentum is a good candidate for the kinetic 4-momentum of a free particle. Indeed, as we will later see, both operators are in the free case connected by a gauge transformation and can thus be physically identified.

\subsection{Electromagnetic interaction}\label{secEM}

We will now derive the Maxwell equations from postulate~\ref{Position} by showing that the commutation relations~(\ref{Px0}) put restrictions on the interaction tensor $\hat F^{\alpha\beta}$ which are equivalent to the Maxwell equations in the presence of an external conserved current, so that $\hat F^{\alpha\beta}$ can be identified with the observable of the electromagnetic field induced by an external current.

Using the commutation relations~\eqref{Px0} and~\eqref{xp0}, we see that the potential 4-momentum commutes with the 4-position,
\begin{eqnarray}
	[\hat A^\alpha,\hat x^\beta]
		&=&[\hat p^\alpha-\hat \wp^\alpha,\hat x^\beta]\\
		&=&[\hat p^\alpha,\hat x^\beta]-[\hat \wp^\alpha,\hat x^\beta]\\
		&=&ig^{\alpha\beta}-ig^{\alpha\beta}=0.
\end{eqnarray}
Together with the fact that both operators act on the same Hilbert space ${\cal E}$ and that $\hat{\mvec x}$ has a non-degenerate spectrum, the condition $[\hat A^\alpha,\hat x^\beta]=0$  implies that $\hat{\mvec A}$ must be diagonal in $\mvec x$, so
\begin{equation}
	\hat{\mvec A}
		=\int d^4x\,\mvec A(\mvec x)|\mvec x\rangle\langle\mvec x|,
\end{equation}
where $\mvec A(\mvec x)$ is a real-valued function on $\mathbbm R^4$. 
Furthermore, we see that the interaction tensor $\hat F^{\alpha\beta}$ defined in~\eqref{PP0} must have the form
\begin{eqnarray}
	\hat F{}^{\alpha\beta}
		&=&i[\hat \wp^\alpha,\hat \wp^\beta]\\
		&=&i[\hat p^\alpha-\hat A^\alpha,\hat p^\beta-\hat A^\beta]\\
		&=&i[\hat p^\alpha,\hat p^\beta]-i[\hat A^\alpha,\hat p^\beta]
		-i[\hat p^\alpha,\hat A^\beta]+i[\hat A^\alpha,\hat A^\beta]\\
		&=&-i[\hat p^\alpha,\hat A^\beta]+i[\hat p^\beta,\hat A^\alpha]\\
		&=&\partial^\alpha A^\beta
		-\partial^\beta A^\alpha,\label{FA}
\end{eqnarray}
where we used that
\begin{equation}
	[\hat A,f(\hat B)]=c\,f'(\hat B)
\end{equation}
for any two operators obeying $[\hat A,\hat B]=c$.
Thus $\hat F^{\alpha\beta}$ is diagonal in $\mvec x$ and its matrix elements are given by 
$
	F^{\alpha\beta}=\partial^\alpha A^\beta
		-\partial^\beta A^\alpha.
$
The interaction tensor is obviously antisymmetric,
\begin{equation}
	\hat F^{\alpha\beta}=-\hat F^{\beta\alpha},
\end{equation}
and self-adjoint,
\begin{equation}
	\hat F^{\alpha\beta}{}^\dag=\hat F^{\alpha\beta}.
\end{equation}
Because $\hat F{}^{\alpha\beta}$ and $\hat{\mvec A}$ are both diagonal in $\mvec x$, they commute,
\begin{equation}
	[\hat F{}^{\alpha\beta},\hat{\mvec A}{}^\gamma]=0,
\end{equation}
and thus
\begin{eqnarray}
	[\hat F{}^{\alpha\beta},\hat \wp^\gamma]
		&=&[\hat F{}^{\alpha\beta},\hat p^\gamma]
		=-i\partial^\gamma F^{\alpha\beta}\label{PF2}.
\end{eqnarray}
Now we use the \emph{Jacobi identity}
\begin{equation}
	[\hat A,[\hat B,\hat C]]+[\hat C,[\hat A,\hat B]]
		+[\hat B,[\hat C,\hat A]]=0
\end{equation}
to see that
\begin{eqnarray}
	[\hat \wp^\gamma,[\hat \wp^\alpha,\hat \wp^\beta]]
	+[\hat \wp^\beta,[\hat \wp^\gamma,\hat \wp^\alpha]]
	+[\hat \wp^\alpha,[\hat \wp^\beta,\hat \wp^\gamma]]=0,
\end{eqnarray}
so together with~(\ref{PP0}) and~(\ref{PF2}) we find
\begin{equation}
	\partial^\gamma F^{\alpha\beta}
		+\partial^\alpha F^{\beta\gamma}
		+\partial^\beta F^{\gamma\alpha}
		=0,\label{maxhom}
\end{equation}
which coincides with the \emph{homogeneous Maxwell equations}. 
Now consider
\begin{equation}
	\partial_\alpha\partial^\alpha\partial_\beta{ A}^\beta
		=\partial_\beta\partial^\beta\partial_\alpha{ A}^\alpha
\end{equation}
and thus
\begin{eqnarray}
	\partial_\alpha\partial^\alpha\partial_\beta{ A}^\beta
		- \partial_\beta\partial^\beta\partial_\alpha{ A}^\alpha
		&=&0\\
	\Leftrightarrow\quad
	\partial_\beta\partial_\alpha\{\partial^\alpha{ A}^\beta
		-\partial^\beta{ A}^\alpha\}&=&0
\end{eqnarray}
which yields together with~\eqref{FA}
\begin{equation}
	\partial_\alpha\partial_\beta F^{\alpha\beta}
		=0.\label{premaxwell}
\end{equation}
If we define the quantity
\begin{equation}\label{calJ}
	{\cal J}^\alpha:=\partial_\beta F^{\alpha\beta},
\end{equation}
then~\eqref{premaxwell} implies that this new quantity fulfills a continuity relation
\begin{equation}
	\partial_\alpha{\cal J}^\alpha=0,
\end{equation}
and can thus be identified with a conserved current of some external source.
Equation~\eqref{calJ} coincides with the inhomogenuous Maxwell equations for a given conserved current.
Definition~\eqref{calJ} can also be written in the more abstract form
\begin{eqnarray}
	\hat{\cal J}^\alpha
		&=&i[\hat \wp_\beta,\hat F^{\beta\alpha}].\label{calJF}
\end{eqnarray}
Now that we have shown that the field tensor $\hat F^{\alpha\beta}$ obeys restrictions which are equivalent to the Maxwell equations in the presence of an external source, we can identify $\hat F^{\alpha\beta}$ with the electromagnetic field tensor and $\hat A^\alpha$ with the electromagnetic 4-potential. 
The electric and magnetic field are obtained from $\hat F^{\alpha\beta}$ by the definitions
\begin{eqnarray}
	\hat E^i&:=&\hat F^{i0}\label{E}\\
	\hat B^i&:=&\frac12\epsilon^{ijk}\hat F^{jk},\label{B}
\end{eqnarray}
which can be inverted as
\begin{eqnarray}
	\hat F^{i0}&=&\hat E^i,\quad
	\hat F^{ij}=\epsilon^{ijk}\hat B^k.
\end{eqnarray}
The electric and magnetic field operators are self-adjoint and diagonal in $\mvec x$ because the field tensor $\hat F^{\alpha\beta}$ is so. They correspond to observables measuring the strength of the electric and magnetic field at each point in spacetime. 
Denote the components of the potential 4-momentum by
\begin{equation}
	\hat{\mvec A}=(\hat  {\sf U},\hat{\rvec A}),
\end{equation}
with $\hat  {\sf U}$ being called the \emph{potential energy} and $\hat{\rvec A}$ being called the \emph{potential momentum}. Then, in spacetime representation the electric and magnetic field~\eqref{E} and~\eqref{B} read
\begin{eqnarray}
	\rvec E&=&-\rvec\nabla  {\sf U}-\frac\partial{\partial t}\rvec A\\
	\rvec B&=&\rvec\nabla\times\rvec A.
\end{eqnarray}
Altogether, the electromagnetic field tensor admits the form
\begin{equation}
	F^{\alpha\beta}=\begin{pmatrix}0&-E^1&-E^2&-E^3\\E^1&0&-B^3&B^2\\
		E^2&B^3&0&-B^1\\E^3&-B^2&B^1&0\end{pmatrix}.
\end{equation}
The homogenous Maxwell equations~(\ref{maxhom}) can be written as
\begin{eqnarray}
	\rvec\nabla\cdot\rvec B&=&0\\
	\rvec\nabla\times\rvec E
		+\frac\partial{\partial t}\rvec B&=&0,
\end{eqnarray}
while the inhomogeneous Maxwell equations~(\ref{calJ}) read
\begin{eqnarray}
	\rvec\nabla\cdot\rvec E&=&{\cal J}^0\\
	\rvec\nabla\times\rvec B
		-\frac\partial{\partial t}\rvec E&=&\rvec{{\cal J}},
\end{eqnarray}
where $\mvec{{\cal J}}$ is the conserved 4-current of an external source.
The vacuum case is a special case where 
\begin{equation}
	\hat{{\cal J}}^\alpha=0,
\end{equation}
which leads together with~\eqref{calJ} to the vacuum Maxwell equations
\begin{equation}
	\partial_\beta F^{\alpha\beta}=0.
\end{equation}
The natural units of electric and magnetic field coincide with those of the electromagnetic field tensor which by definition~\eqref{PP0} are given by
\begin{equation}
	[F^{\alpha\beta}]=[p^\alpha][p^\beta]=\frac1{{\rm m}^2}.
\end{equation}

\subsection{Quantum processes}

In standard quantum mechanics, a particle state is defined by its properties, and these properties are time-dependent: As time goes by, the particle changes its properties and therefore changes its state. In QET, a quantum event is defined by its properties, but these properties are not time-dependent: An event \emph{occurs} and does not \emph{evolve}.
So how can the evolution of a particle actually be described in this framework, or in other words: Where is the dynamics?

The basic ingredient for a dynamical theory is the causation between individual events.
Let us define that two events $\psi$ and $\phi$ where the first one causes the second, constitute an elementary \emph{quantum process} written as $\psi\rightarrow\phi$. The task of the theory is now to give us the \emph{probability} for any such process.
\jbox{
\begin{Postulate}[Quantum Process]\label{Evolution}
Any observable process involving a spinless particle is a discrete jump process in the event Hilbert space ${\cal E}=L^2(\mathbbm R^4)$. The \emph{transition amplitude} for the process $\psi\rightarrow\phi$ is defined by the matrix elements
\begin{equation}\label{TransitA}
	\tau(\phi,\psi):=\langle\phi|\hat{\cal G}|\psi\rangle
\end{equation}
where the \emph{propagator} $\hat{\cal G}$ is defined by
\begin{equation}\label{propagator}
	\hat{\cal G}:=\delta(\hat{\mvec\wp}{}^2-m^2).
\end{equation}
An event $\psi\in{\cal E}$ is called \emph{physically allowed} if its self-amplitude does not vanish, that is
\begin{equation}
	\tau(\psi,\psi)\neq0.
\end{equation}
The probability for the process $\psi\rightarrow\phi$ is for any two physically allowed events $\psi,\phi\in{\cal E}$ given by
\begin{eqnarray}\label{TransitP}
	P(\phi,\psi)&:=&\frac{|\tau(\phi,\psi)|^2}
		{\tau(\phi,\phi)\tau(\psi,\psi)}.
\end{eqnarray}
\end{Postulate}
}
The propagator $\hat{\cal G}$ obviously takes care that the event is located on the kinetic mass shell and does not leave it while undergoing a transition. 
If we define the operator for the \emph{kinetic mass square} via
\begin{equation}
	\hat m^2 := \hat{\mvec\wp}{}^2,
\end{equation}
then postulate \ref{Evolution} is the mathematical correlate of the following physical principle:
\begin{quote}\em
Any physically allowed quantum process takes place on the kinetic mass shell corresponding to a fixed mass parameter which is specific to the type of the particle under study.
\end{quote}
Note that here and in the following we understand the kinetic mass shell as consisting of the upper and the lower part corresponding to positive and negative kinetic energy, respectively. Later we will derive the familiar physical interpretation in terms of particles and antiparticles.
 
The transition amplitude $\tau$ is a sesquilinear form, that is,
\begin{equation}\label{sesqui}
	\begin{split}
	\tau(a\phi,b\psi)&=a^*b\,\tau(\phi,\psi)
	,\quad a,b\in\mathbbm C\\
	\tau(\phi_1+\phi_2,\psi)&=\tau(\phi_1,\psi)+\tau(\phi_2,\psi)\\
	\tau(\phi,\psi_1+\psi_2)&=\tau(\phi,\psi_1)+\tau(\phi,\psi_2).
	\end{split}
\end{equation}
Furthermore we have
\begin{equation}
	\tau(\psi,\phi)=\tau^*(\phi,\psi)
\end{equation}
and
\begin{equation}\label{Ageq0}
	\tau(\psi,\psi)\geq0,
\end{equation}
because the propagator $\hat{\cal G}$ is symmetric and positive, which we may denote a bit sloppily as
\begin{equation}
	\hat{\cal G}^\dag=\hat{\cal G},\quad\hat{\cal G}\geq0.
\end{equation}
Altogether, the transition amplitude $\tau$ is a positive semidefinite hermitian form on ${\cal E}$.
Now let $\lambda$ be any real number, then because of~\eqref{Ageq0} we have
\begin{eqnarray}
	0&\leq&\tau(\psi+\lambda\phi,\psi+\lambda\phi)\\
		&=&
		\tau(\psi,\psi)+2\lambda{\rm Re}\{\tau(\phi,\psi)\}
		+\lambda^2\tau(\phi,\phi)\\
		&\leq&\tau(\psi,\psi)+2\lambda|\tau(\phi,\psi)|
		+\lambda^2\tau(\phi,\phi)
\end{eqnarray}
The righthand side of this inequality is a quadratic function in $\lambda$ of the form $f(\lambda)=a\lambda^2+b\lambda+c$, which is only nonnegative if the coefficients fulfill $b^2\leq 4ac$, thus
\begin{equation}\label{alphaineq}
	|\tau(\phi,\psi)|^2\leq\tau(\phi,\phi)\tau(\psi,\psi).
\end{equation}
By~\eqref{Ageq0} and \eqref{alphaineq} we have
\begin{equation}\label{POK}
	0\leq P(\phi,\psi)\leq1,
\end{equation}
as required for a probability interpretation.
The transition amplitude can be written as
\begin{equation}
	\tau(\phi,\psi)=\int d^4x\int d^4x'\,\phi^*(\mvec x)
		{\cal G}(\mvec x,\mvec x')\psi(\mvec x),
\end{equation}
where 
\begin{equation}
	{\cal G}(\mvec x,\mvec x')=\langle\mvec x|\hat{\cal G}|\mvec x'\rangle
\end{equation}
is the \emph{propagator function}.
The propagator $\hat{\cal G}$ takes the role of the unitary time evolution operator $\hat U(t,t')$ in standard quantum mechanics.
In terms of the propagator, the transition probabilities~\eqref{TransitP} are given by
\begin{equation}\label{TransitPG}
	P(\phi,\psi)=\frac{|\langle\phi|\hat{\cal G}|\psi\rangle|^2}
		{\langle\phi|\hat{\cal G}|\phi\rangle\langle\psi|\hat{\cal G}|\psi\rangle}.
\end{equation}
These probabilities are defined for all pairs of physically allowed events, which do not form a mutually exclusive and exhaustive set. They can, however, be interpreted in the following sense. If in $N_\psi$ independent and identical experiments the event $|\psi\rangle$ occurs and there is a detector for the occurence of another event $|\phi\rangle$, then the number $M_\phi$ of detector clicks corresponding to $\phi$ approximates
\begin{equation}
	M_\phi\rightarrow P(\phi,\psi)N_\psi
\end{equation}
in the limit of $N_\psi\rightarrow\infty$. The number $P(\phi,\psi)$ can be regarded as the tendency that the event $\psi$ causes the event $\phi$. And because $P(\phi,\psi)$ is symmetric, the same tendency holds for the other causal direction. Such a notion of probability is certainly more rudimentary than the usual one, which applies to an exclusive and exhaustive set of individual events.

\section{Transformations}

\subsection{Translation}

Let us see how total energy and momentum arise naturally as the generators of spacetime translations.
For any $\mvec a\in\mathbbm R^4$ the spacetime translation $T({\mvec a})\in{\cal T}$ from the translation group ${\cal T}$ acts on an event $\mvec x\in\mathbbm R^4$ as
\begin{equation}
	T({\mvec a}):\quad \mvec x\mapsto \mvec x+\mvec a.
\end{equation}
By applying the quantization rule~\eqref{quant} we find that the corresponding quantum operation is given by
\begin{equation}
	\hat U_T(\mvec a):\quad |\mvec x\rangle\mapsto|\mvec x+\mvec a\rangle,
\end{equation}
so the translation operator can be written as
\begin{equation}\label{UT1}
	\hat U_T(\mvec a)=\int d^4x\,|\mvec x+\mvec a\rangle\langle\mvec x|
\end{equation}
which is a unitary operator on ${\cal E}$, as can easily be verified. Therefore the translation group ${\cal T}$ has a unitary representation
\begin{equation}
	{\cal U}_T:=\{\hat U_T(\mvec a)\mid \mvec a\in\mathbbm R^4\}
\end{equation}
on the event space ${\cal E}$. The translations form a Lie group, so let us look for the generators. In spacetime representation, the action of $\hat U_T(\mvec a)$ on an event $|\psi\rangle\in{\cal E}$ reads 
\begin{eqnarray}
	\hat U_T(\mvec a)\psi(\mvec x)
		&=&\langle\mvec x|\hat U_T(\mvec a)|\psi\rangle\\
		&=&\langle\mvec x-\mvec a|\psi\rangle\\
		&=&\psi(\mvec x-\mvec a).
\end{eqnarray}
For an infinitesimal translation $T(d\mvec a)$ we have
\begin{eqnarray}
	\hat U_T(d\mvec a)\psi(\mvec x)&=&\psi(\mvec x-d\mvec a)\\
		&=&\psi(\mvec x)-da^\alpha\frac{\partial\psi(\mvec x)}{\partial x^\alpha}\\
		&=&\psi(\mvec x)-(d\mvec a\cdot\mvec\partial)\psi(\mvec x)\\
		&=&\psi(\mvec x)+i\,(d\mvec a\cdot\hat{\mvec p})\psi(\mvec x),
\end{eqnarray}
where 
\begin{equation}
	\mvec\partial:=\frac\partial{\partial\mvec x}
		=(\frac\partial{\partial t},-{\rvec\nabla_x})
\end{equation}
is the \emph{4-derivative} and where we have used definition~\eqref{p} for the \emph{total 4-momentum operator}, which is thus the generator of the group ${\cal U}_T$.
The finite transformation is thus given by
\begin{equation}\label{UT2}
	\hat U_T(\mvec a)=e^{i\mvec a\hat{\mvec p}}.
\end{equation}
Let us find the eigenvectors of $\hat{\mvec p}$. For any eigenvector $|\mvec p\rangle$ we have 
\begin{equation}
	\langle\mvec x|\hat{\mvec p}|\mvec p\rangle
		=\mvec p\langle\mvec x|\mvec p\rangle,
\end{equation}
and also
\begin{equation}
	\langle\mvec x|\hat{\mvec p}|\mvec p\rangle
		=i\mvec\partial\langle\mvec x|\mvec p\rangle.
\end{equation}
This implies the differential equation
\begin{equation}
	i\mvec\partial\langle\mvec x|\mvec p\rangle
		=\mvec p\langle\mvec x|\mvec p\rangle,
\end{equation}
which is solved by
\begin{equation}\label{xpbracket}
	\langle\mvec x|\mvec p\rangle= \frac1{\sqrt{2\pi}^4}\,e^{-i\mvec p\mvec x},
\end{equation}
where we have readily chosen the normalization constant such that the orthonormality condition
\begin{eqnarray}
	\langle\mvec p|\mvec p'\rangle&=&\delta(\mvec p-\mvec p')
\end{eqnarray}
is fulfilled.
Applying the completeness relation~\eqref{xcomplete} to $|\mvec p\rangle$ and using~\eqref{xpbracket}, we find that the eigenvectors $|\mvec p\rangle$ are obtained by Fourier transformation of the position eigenvectors $|\mvec x\rangle$,
\begin{equation}
	|\mvec p\rangle=\frac1{\sqrt{2\pi}^4}\int d^4x\,
	e^{-i\mvec p\mvec x}|\mvec x\rangle,
\end{equation}
from where we find that
\begin{equation}
	\int d^4p\,|\mvec p\rangle\langle\mvec p|=\mathbbm1.
\end{equation}
Thus the set
\begin{equation}
	{\cal B}_p:=\{|\mvec p\rangle\mid\mvec p\in\mathbbm R^4\}
\end{equation}
constitutes an orthonormal and complete basis of ${\cal E}$.
A 4-momentum eigenvector is transformed under translation as
\begin{equation}
	\hat U_T(\mvec a)|\mvec p\rangle=e^{i\mvec p\mvec a}|\mvec p\rangle.
\end{equation}
Let us use the shorthand notation
\begin{equation}
	\mvec p\equiv (E,\rvec p).
\end{equation}
where $E$ is the total energy and $\rvec p$ is the total momentum.
The energy operator $\hat E$ thus appears as the generator of time translations, while the momentum operator $\hat{\rvec p}$ appears as the generator of space translations. Since they are self-adjoint operators on ${\cal E}$ they represent properties.
In spacetime representation, energy and momentum obtain in the form
\begin{equation}
	\hat E=i\frac\partial{\partial t},\quad
	\hat{\rvec p}=-i\rvec\nabla_x.
\end{equation}
As one can easily verify, the 4-position is affected by a translation while the 4-momentum is not:
\begin{eqnarray}
	\hat U_T^\dag(\mvec a)\hat{\mvec x}\hat U_T(\mvec a)
		&=&\hat{\mvec x}+\mvec a,\\
	\hat U_T^\dag(\mvec a)\hat{\mvec p}\hat U_T(\mvec a)
		&=&\hat{\mvec p}.
\end{eqnarray}
The improper vector $|\mvec p\rangle=|E,\rvec p\rangle$ thus represents an event with energy $E$ and momentum $\rvec p$. It is impossible to specify when and where this event occurs, because $|\mvec p\rangle$ has an infinite uncertainty in time and space, because its amplitude is a plain wave in spacetime. 

Let us fix the natural unit of the 4-momentum. Because $\mvec p\mvec x$ is in the exponent it must be free of units, each component of $\mvec p$ should have the unit of a reverse position:
\begin{equation}
	[p^\alpha]=\frac1{[x^\alpha]}=\frac1{\rm m}.
\end{equation}
By construction then, the Planck constant is $\hbar=1$.

\subsection{Lorentz transformation}

The group ${\cal L}$ of Lorentz transformations enters the stage when the observer is taken to a different reference frame.
If an event has the coordinates $\mvec x$ within the original reference frame then in another reference frame its coordinates are Lorentz transformed by the matrix $\Lambda\in{\cal L}$ according to
\begin{equation}\label{lorentz}
	\Lambda: \mvec x\mapsto\Lambda \mvec x.
\end{equation}
The Lorentz group consists of all $4\times4$-matrices $\Lambda$ that leave scalar products $\mvec x\mvec y$ invariant:
\begin{equation}
	(\Lambda\mvec x)(\Lambda\mvec y)=\mvec x\mvec y,\quad
		\forall \mvec x,\mvec y\in\mathbbm R^4,
\end{equation}
where the scalar product is defined by
\begin{equation}
	\mvec x\mvec y:=x_\alpha y^\alpha\equiv g_{\alpha\beta}x^\alpha y^\beta
\end{equation}
and where the Einstein summation convention is used and the metric tensor of special relativity is defined by
\begin{equation}
	g_{\mu\nu}=\begin{pmatrix}1&0&0&0\\0&-1&0&0\\0&0&-1&0\\0&0&0&-1\end{pmatrix}.
\end{equation}
Lorentz invariance of scalars can also be expressed as
\begin{equation}
	\mvec x(\Lambda\mvec y)=(\Lambda^{-1}\mvec x)\mvec y,\quad
		\forall \mvec x,\mvec y\in\mathbbm R^4.
\end{equation}
Applying the quantization rule~\eqref{quant}, equation~\eqref{lorentz} implies that the corresponding operator $\hat U_\Lambda$ is defined by
\begin{equation}
	\hat U_\Lambda: |\mvec x\rangle\mapsto|\Lambda\mvec x\rangle,
\end{equation}
and can thus be written as
\begin{equation}\label{ULambda1}
	\hat U_\Lambda=\int d^4x\,|\Lambda\mvec x\rangle\langle\mvec x|.
\end{equation}
Let us show that this defines a unitary transformation:
\begin{eqnarray}
	\hat U_\Lambda^\dag\hat U_\Lambda&=&\int d^4x\int d^4x'\,
		|\mvec x\rangle\langle\Lambda\mvec x|\Lambda\mvec x'\rangle
		\langle\mvec x'|\\
		&=&\int d^4x\int d^4x'\,\delta(\Lambda\mvec x-\Lambda\mvec x')
		|\mvec x\rangle\langle\mvec x'|\\
		&=&\int d^4x\int d^4x'\,\frac1{|\det\Lambda|}
		\delta(\mvec x-\mvec x')
		|\mvec x\rangle\langle\mvec x'|\\
		&=&\int d^4x\,|\mvec x\rangle\langle\mvec x|=\mathbbm 1,
\end{eqnarray}
where we have used that $\det\Lambda=\pm1$ for all $\Lambda\in{\cal L}$.
The unitary operator $\hat U_\Lambda$ acts on $|\mvec p\rangle$ as
\begin{eqnarray}
	\hat U_\Lambda|\mvec p\rangle
		&=&\hat U_\Lambda\frac1{\sqrt{2\pi\hbar}^4}\int d^4x\,
		e^{-\frac i\hbar\mvec p\mvec x}|\mvec x\rangle\\
		&=&\frac1{\sqrt{2\pi\hbar}^4}\int d^4x\,
		e^{-\frac i\hbar\mvec p\mvec x}|\Lambda\mvec x\rangle\\
		&=&\frac1{\sqrt{2\pi\hbar}^4}\int d^4x\,
		e^{-\frac i\hbar\mvec p(\Lambda^{-1}\mvec x)}|\mvec x\rangle\\
		&=&\frac1{\sqrt{2\pi\hbar}^4}\int d^4x\,
		e^{-\frac i\hbar(\Lambda\mvec p)\mvec x}|\mvec x\rangle\\
		&=&|\Lambda\mvec p\rangle,
\end{eqnarray}
therefore the operators $\hat{\mvec x}$ and $\hat{\mvec p}$ are correctly transformed under the action of $\hat U_\Lambda$,
\begin{eqnarray}
	\hat U_\Lambda^\dag\hat{\mvec x}\hat U_\Lambda&=&\Lambda\hat{\mvec x},\\
	\hat U_\Lambda^\dag\hat{\mvec p}\hat U_\Lambda&=&\Lambda\hat{\mvec p},
\end{eqnarray}
and can be regarded as \emph{vector operators}.
In the spacetime representation $\hat U_\Lambda$ acts on event wave functions $\psi(\mvec x)\in L^2(\mathbbm R^4)$ as
\begin{eqnarray}
	\hat U_\Lambda\psi(\mvec x)&=&\langle\mvec x|\hat U_\Lambda|\psi\rangle
		=\langle\Lambda^{-1}\mvec x|\psi\rangle
		=\psi(\Lambda^{-1}\mvec x).
\end{eqnarray}
Concluding, the Lorentz group ${\cal L}$ has a unitary representation 
\begin{equation}
	{\cal U}_{\cal L}:=\{\hat U_\Lambda\mid \Lambda\in{\cal L}\}
\end{equation}
on the event space ${\cal E}$.

\subsection{Rotation and boost}

The subgroup ${\cal L}_+^\uparrow\subset{\cal L}$ of proper orthochronuous  Lorentz transformations is the group of rotations and boosts. Here, angular momentum and boost operator are the generators of rotation and boost, respectively.

The group  ${\cal L}_+^\uparrow$ is a Lie group that can be parametrized by six parameters. The first three parameters, written as $\rvec\varphi=(\varphi_1,\varphi_2,\varphi_3)$, correspond to a rotation about the axis $\rvec n$ by the angle $\varphi$, so that $\rvec\varphi=\varphi\rvec n$. The last three parameters, written as $\rvec\theta=(\theta_1,\theta_2,\theta_3)$, correspond to a boost in the direction of $\rvec m$ by the rapidity $\theta$, so that $\rvec\theta=\theta\rvec m$. The rapidity $\theta$ is defined by $\theta={\rm artanh}(v)$ where $v$ is the velocity of the boost.
One can combine these six parameters into one antisymmetric $4\times 4$-matrix
$\omega$ 
so that
\begin{eqnarray}
	\omega_{\alpha\beta}&=&-\omega_{\beta\alpha}\\
	\omega_{0a}&=&\theta_a\\
	\omega_{ij}&=&\epsilon_{ijk}\varphi_c.
\end{eqnarray}
An infinitesimal Lorentz tranformation of the event $\mvec x$ then reads
\begin{eqnarray}
	\Lambda\mvec x&=&\mvec x+d\omega\mvec x\\
	\text{i.e. }\quad
	\Lambda^\alpha{}_\beta x^\beta&=&x^\alpha+d\omega^\alpha{}_\beta x^\beta.
\end{eqnarray}
This implies that the action of an infinitesimal $\hat U_\Lambda$ on a quantum event $|\psi\rangle$ would read in spacetime representation
\begin{eqnarray}
	\hat U_\Lambda\psi(\mvec x)
		&=&\psi(\mvec x-d\omega\mvec x)\\
		&=&\psi(\mvec x)-d\omega^\alpha{}_\beta x^\beta
		\partial_\alpha\psi(\mvec x)\\
		&=&\psi(\mvec x)-i\,d\omega_{\alpha\beta} 
		\hat x^\alpha \hat p^\beta\psi(\mvec x).
\end{eqnarray}
Because of the antisymmetry of $\omega$ we can write
\begin{eqnarray}
	d\omega_{\alpha\beta}\hat x^\alpha \hat p^\beta
		&=&\frac12\,d\omega_{\alpha\beta}(\hat x^\alpha\hat p^\beta
		-\hat x^\beta \hat p^\alpha)\\
		&=&\frac12\,d\omega_{\alpha\beta}\hat M^{\alpha\beta},
\end{eqnarray}
where 
\begin{equation}
	\hat M^{\alpha\beta}:=\hat x^\alpha\hat p^\beta-\hat x^\beta\hat p^\alpha
\end{equation}
is hence the generator of the group. 
As a consequence of the commutator relation~\eqref{xp0}, the generator $\hat M$ is a self-adjoint operator:
\begin{eqnarray}
	(\hat M^{\alpha\beta})^\dag
		&=&\hat p^\beta\hat x^\alpha-\hat p^\alpha\hat x^\beta\\
		&=&\hat x^\alpha\hat p^\beta+ig^{\beta\alpha}
		-(\hat x^\beta\hat p^\alpha+ig^{\alpha\beta})\\
		&=&\hat x^\alpha\hat p^\beta-\hat x^\beta\hat p^\alpha\\
		&=&\hat M^{\alpha\beta}.
\end{eqnarray}
Consequently, a finite proper orthochronuous Lorentz transformation is represented by the unitary operator
\begin{equation}
	\hat U_\Lambda(\omega)=e^{-\frac i2\omega_{\alpha\beta}\hat M^{\alpha\beta}}.
\end{equation}
Defining the \emph{angular momentum operator}
\begin{eqnarray}
	\hat L^i &:=&\frac12 \epsilon^{ijk}\hat M^{bc}
		=\epsilon^{ijk}\hat x^j\hat p^k\\
	\text{i.e.}\quad
	\hat{\rvec L}&=&\hat{\rvec x}\times\hat{\rvec p}
\end{eqnarray}
and the \emph{boost operator}
\begin{eqnarray}
	\hat K^i&:=&\hat M^{0a}
		=\hat x^0\hat p^i-\hat x^i\hat p^0\\
	\text{i.e.}\quad
	\hat{\rvec K}&=&\hat t\hat{\rvec p}-\hat{\rvec x}\hat E
\end{eqnarray}
we arrive at the representation
\begin{equation}\label{ULambda2}
	\hat U_\Lambda(\rvec\varphi,\rvec\theta)
		=e^{-i\rvec\varphi\hat{\rvec L}-i\rvec\theta\hat{\rvec K}}.
\end{equation}
Using the commutation relation~\eqref{xp0} one can verify that the operators $\hat{\rvec L}$ and $\hat{\rvec K}$ obey the commutation relations
\begin{eqnarray}
	[\hat L^i,\hat L^j]&=&i\epsilon^{ijk}\hat L^k\\{}
	[\hat L^i,\hat K^j]&=&i\epsilon^{ijk}\hat K^k\\{}
	[\hat K^i,\hat K^j]&=&-i\epsilon^{ijk}\hat L^k.
\end{eqnarray}
As one can see by looking at~\eqref{ULambda2}, the angular momentum operator $\hat{\rvec L}$ is the generator of rotations, while the boost operator $\hat{\rvec K}$ is the generator of boosts.
Because both operators are self-adjoint, they represent observable qualities of an event. Let us fix the natural units of $\rvec L$ and $\rvec K$. By their definition we obtain
\begin{eqnarray}
	[L^i]=[K^i]&=&[x^i][p^i]={\rm m}\cdot\frac 1{\rm m}=1,
\end{eqnarray}
thus they are dimensionless.

The proper orthochronuous Lorentz group ${\cal L}_+^\uparrow$ has the unitary representation
\begin{equation}
	{\cal U}_{{\cal L}_+^\uparrow}=\{\hat U_\Lambda(\rvec\varphi,\rvec\theta)
		\mid\rvec\varphi\in[-\pi,\pi]^3,\rvec\theta\in\mathbbm R^3\}
\end{equation}
on ${\cal E}$.
The 4-position and the 4-momentum are transformed as
\begin{eqnarray}
	\hat U_\Lambda^\dag\hat{\mvec x}\hat U_\Lambda
		&=&\Lambda\hat{\mvec x}\\
	\hat U_\Lambda^\dag\hat{\mvec p}\hat U_\Lambda
		&=&\Lambda\hat{\mvec p}.
\end{eqnarray}

\subsection{Parity and time reversal}\label{secPT}

Now we come to the discrete transformations.
The \emph{parity transformation} $P$ flips the sign of the spacelike coordinates, 
\begin{equation}
	P: (t,\rvec x)\mapsto (t,-\rvec x).
\end{equation}
The quantization rule~\eqref{quant} yields the corresponding unitary operator
\begin{equation}
	\hat{\cal P}=\int d^4x\,|t,-\rvec x\rangle\langle t,\rvec x|.
\end{equation}
The parity transformation operator is self-adjoint,
\begin{equation}
	\hat{\cal P}^\dag=\hat{\cal P},
\end{equation}
thus it corresponds to an observable quality of an event, called the \emph{parity}. Because $\hat{\cal P}^2=\mathbbm 1$ the eigenvalues of $\hat{\cal P}$ are $\pm 1$, corresponding to \emph{pair} and \emph{impair} events, respectively.
The parity transformation sends the position and momentum operator to their negative,
\begin{eqnarray}
	\hat {\cal P}\hat{\rvec x}\hat {\cal P}&=&-\hat{\rvec x}\\
	\hat {\cal P}\hat{\rvec p}\hat {\cal P}&=&-\hat{\rvec p}.
\end{eqnarray}
The \emph{time reversal transformation} flips the sign of the timelike component,
\begin{equation}
	T: (t,\rvec x)\mapsto (-t,\rvec x).
\end{equation}
Following the quantization rule~\eqref{quant}, the corresponding unitary operator reads
\begin{equation}
	\hat{\cal T}=\int d^4x\,|\!-\!t,\rvec x\rangle\langle t,\rvec x|.
\end{equation}
The time reversal operator is self-adjoint,
\begin{equation}
	\hat{\cal T}^\dag=\hat{\cal T},
\end{equation}
thus it represents an observable quality of an event which we shall call the \emph{reversibility}. Because $\hat{\cal T}^2=\mathbbm1$, the eigenvalues of $\hat{\cal T}$ are $\pm 1$ corresponding to \emph{reversible} and \emph{antireversible} events, respectively.
The time reversal transformation sends the time and energy operator to their negative,
\begin{eqnarray}
	\hat{\cal T}\hat t\hat{\cal T}&=&-\hat t\\
	\hat{\cal T}\hat E\hat{\cal T}&=&-\hat E.
\end{eqnarray}
As the parity and time reversal transformation are discrete transformations, they have no generators. But since they are self-adjoint operators, they represent properties themselves.
The natural units of ${\cal P}$ and ${\cal T}$ are
\begin{equation}
	[{\cal P}]=[{\cal T}]=1.
\end{equation}
Under parity transformation and time reversal, the 4-position and the 4-momentum are transformed as
\begin{eqnarray}
	\hat{\cal P}\hat{\cal T}\hat{\mvec x}\hat{\cal P}\hat{\cal T}
		&=&-\hat{\mvec x}\\
	\hat{\cal P}\hat{\cal T}\hat{\mvec p}\hat{\cal P}\hat{\cal T}
		&=&-\hat{\mvec p}.
\end{eqnarray}
The full Lorentz group ${\cal L}$ is obtained by joining
\begin{equation}
	{\cal L}={\cal L}_+^\uparrow\cup P{\cal L}_+^\uparrow
		\cup T{\cal L}_+^\uparrow\cup PT{\cal L}_+^\uparrow,
\end{equation}
which is unitarily represented by
\begin{equation}
	{\cal U}_{\cal L}={\cal U}_{{\cal L}_+^\uparrow}
		\cup \hat{\cal P}{\cal U}_{{\cal L}_+^\uparrow}
		\cup \hat{\cal T}{\cal U}_{{\cal L}_+^\uparrow}
		\cup \hat{\cal P}\hat{\cal T}{\cal U}_{{\cal L}_+^\uparrow}.
\end{equation}

\subsection{Poincar\'e transformation}

Now we join the Lorentz group ${\cal L}$ with the translation group ${\cal T}$ to obtain the full \emph{Poincar\'e group}
\begin{equation}
	{\cal P}:={\cal L}\cup{\cal T}. 
\end{equation} 
An event $\mvec x\in\mathcal R^4$ is Poincar\'e transformed by the Lorentz matrix $\Lambda\in{\cal L}$ and the translation $T(\mvec a)\in{\cal T}$ as
\begin{equation}
	L(\Lambda,\mvec a): \mvec x\mapsto \Lambda\mvec x+\mvec a.
\end{equation}
According to~\eqref{quant} on the event space ${\cal E}$ such Poincar\'e transformation is represented by the unitary operator
\begin{equation}
	\hat U_L(\Lambda,\mvec a)|\mvec x\rangle
		=|\Lambda\mvec x+\mvec a\rangle
\end{equation}
or
\begin{equation}
	\hat U_L(\Lambda,\mvec a)=\int d^4x\,
		|\Lambda\mvec x+\mvec a\rangle\langle\mvec x|.
\end{equation}
The above unitary operation can be decomposed into
\begin{equation}\label{Udecomp}
	\hat U_L(\Lambda,\mvec a)=\hat U_T(\mvec a)\hat U_\Lambda,
\end{equation}
where the translation operator is given by~(\ref{UT1}) and the Lorentz operator by~\eqref{ULambda1}.
Since translation and Lorentz transformation do not commute, the ordering of the operators in~\eqref{Udecomp} is important. Concluding, the Poincar\'e group ${\cal P}$ has a unitary representation
\begin{equation}
	{\cal U}_{\cal P}=\{\hat U_L\mid L\in{\cal P}\}
\end{equation}
on the event space ${\cal E}$. The eigenvectors of the 4-momentum $\hat{\mvec p}$ are transformed as
\begin{equation}
	\hat U_L(\Lambda,\mvec a)|\mvec p\rangle
		=e^{i\Lambda\mvec p\mvec a}|\Lambda\mvec p\rangle.
\end{equation}
The 4-position and the 4-momentum operator are transformed as
\begin{eqnarray}
	\hat U_L^\dag(\Lambda,\mvec a)\,\hat{\mvec x}\,\hat U_L(\Lambda,\mvec a)
		&=&\Lambda\hat{\mvec x}+\mvec a\label{ULx}\\
		\hat U_L^\dag(\Lambda,\mvec a)\,\hat{\mvec p}\,\hat U_L(\Lambda,\mvec a)
		&=&\Lambda\hat{\mvec p}.\label{ULp}
\end{eqnarray}

\subsection{Poincar\'e invariance}

The following theorem shows that identical experiments performed in different reference frames yield identical results, hence QET obeys the principle of special relativity.
\jbox{
\begin{Theorem}[Poincar\'e invariance]\label{TheoPoin}
For any Poincar\'e transformation $L(\Lambda,\mvec a)\in{\cal P}$ the transition amplitudes~\eqref{TransitA} are invariant under the transformation
\begin{eqnarray}
	|\psi\rangle
		&\mapsto&|\psi'\rangle
		=\hat U_L(\Lambda,\mvec a)|\psi\rangle\label{PoinTrafoPsi}\\
	\hat{\mvec A}
		&\mapsto&\hat{\mvec A}'
		=\Lambda\hat{\mvec A}(\Lambda^{-1}(\hat{\mvec x}-\mvec a)),
		\label{PoinTrafoA}
\end{eqnarray}
where $\hat U_L(\Lambda,\mvec a)\in{\cal U}_{\cal P}$ is the unitary representation of $L(\Lambda,\mvec a)$ on ${\cal E}$, where $|\psi\rangle\in{\cal E}$ is any quantum event and where $\mvec A$ is the 4-potential.
\end{Theorem}
}
\emph{Proof}.
Using~\eqref{ULx} one finds that 
\begin{eqnarray}
	\hat U_L^\dag(\Lambda,\mvec a)\mvec A(\hat{\mvec x})\hat U_L(\Lambda,\mvec a)
		&=&\mvec A(\Lambda\hat{\mvec x}+\mvec a).
\end{eqnarray}
Therefore, by further use of~\eqref{ULp}, the transition amplitudes are Poincar\'e transformed as
\begin{eqnarray}
	\tau(\phi,\psi)
		&=&\langle\phi|\delta(\hat{\mvec\wp}{}^2-m^2)|\psi\rangle\\
		&=&\langle\phi|\delta((\hat{\mvec p}{}-\hat{\mvec A})^2-m^2)|\psi\rangle\\
		&\mapsto&\langle\phi|\hat U_L^\dag
		\delta\big((\hat{\mvec p}-\Lambda{\mvec A}
		(\Lambda^{-1}(\hat{\mvec x}-\mvec a)))^2-m^2\big)\hat U_L
		|\psi\rangle\\
		&=&\langle\phi|\delta\big((\Lambda\hat{\mvec p}
		-\Lambda\mvec A(\hat{\mvec x}))^2-m^2\big)
		|\psi\rangle\\
		&=&\langle\phi|\delta\big((\hat{\mvec p}
		-\hat{\mvec A}{})^2-m^2\big)|\psi\rangle\\
		&=&\tau(\phi,\psi),
\end{eqnarray}
which completes the proof of the theorem.\hfill$\Box$

\subsection{Charge conjugation}

Let us define the family of amplitudes for the transition of a quantum event from $\psi\in{\cal E}$ to the spacetime points $\mvec x\in\mathbbm R^4$ as the \emph{event orbit} corresponding to the initial event $\psi$, and denote it by the corresponding capital letter $\Psi$, so 
\begin{equation}\label{orbit}
	\Psi(\mvec x):=\langle\mvec x|\hat{\cal G}|\psi\rangle.
\end{equation}
The orbit then has the form
\begin{equation}
	\Psi(\mvec x)=\int d^4x\,{\cal G}(\mvec x,\mvec x')\psi(\mvec x'),
\end{equation}
which resembles to the propagated wave function in standard quantum mechanics, except that the integration is carried out over $\mathbbm R^4$ instead of $\mathbbm R^3$, which is good because it is manifest covariant.
In standard QM, the wave function $\Psi(t,\rvec x)$ represents the amplitude to find the particle at time $t$ and at position $\rvec x$. In analogy to that, in QET the orbit function $\Psi(t,\rvec x)$ represents the amplitude for the transition $\psi\rightarrow(t,\rvec x)$.
Because the propagator is given by $\hat{\cal G}=\delta(\hat{\wp}^2-m^2)$, we find that
\begin{equation}
	(\hat{\wp}^2-m^2)\hat{\cal G}=0,
\end{equation}
and thus
\begin{eqnarray}
	0&=&\langle\mvec x|(\hat{\wp}^2-m^2)\hat{\cal G}|\psi\rangle\\
		&=&\left((i\mvec\partial-\mvec A(\mvec x))^2-m^2\right)
		\langle\mvec x|\hat{\cal G}|\psi\rangle,
\end{eqnarray}
therefore the orbit~\eqref{orbit} obeys the equation
\begin{equation}\label{orbiteq}
	\left((i\mvec\partial-\mvec A(\mvec x))^2-m^2\right)\Psi(\mvec x)=0.
\end{equation}
In section~\ref{secEM}, $\mvec A$ has been identified with the electromagnetic 4-potential, so the orbit $\Psi(\mvec x)$ obeying equation~\eqref{orbiteq} describes a spinless particle exposed to the electromagnetic potential $\mvec A$.
Taking the complex conjugate of the above equation we obtain
\begin{equation}
	\left((i\mvec\partial+\mvec A(\mvec x))^2-m^2\right)\Psi^*(\mvec x)=0.
\end{equation}
Hence, the event orbit $\Psi^*(\mvec x)$ describes a particle of \emph{opposite charge} exposed to the electromagnetic potential $\mvec A$. Therefore the transformation
\begin{equation}
	\Psi(\mvec x)\mapsto \Psi^*(\mvec x)
\end{equation}
realizes a \emph{charge conjugation}. Since $\Psi(\mvec x)$ represents the amplitude for the transition $\psi\rightarrow\mvec x$, the complex conjugate orbit 
\begin{equation}
	\Psi^*(\mvec x)=\langle\psi|\hat{\cal G}|\mvec x\rangle
\end{equation}
represents the amplitude for the \emph{reversed transition} $\mvec x\rightarrow\psi$. 
Because the eigenvectors $|\mvec x\rangle$ form a complete basis of ${\cal E}$ we can deduce the following theorem:
\jbox{
\begin{Theorem}[Charge conjugation]\label{TheoCharge}
The propagation of a particle from $\psi$ to $\phi$ is identical to the propagation of a particle of opposite charge from $\phi$ to $\psi$.
\end{Theorem}
}
A particle having opposite charge but with all other properties remaining unchanged is identified with an \emph{antiparticle}.
If we denote the propagation of an antiparticle from $\psi$ to $\phi$ by $\psi\stackrel-\rightarrow\phi$, then we can write down theorem~\ref{TheoCharge} in the compact form
\begin{equation}
	\psi\stackrel-\rightarrow\phi = \psi\leftarrow\phi.
\end{equation}
In particular, a particle propagating backwards in time and space is indistinguishable from an antiparticle propagating forwards in time and space, or
\begin{equation}
	(t,\rvec x)\stackrel-\rightarrow(t',\rvec x')
		=(t,\rvec x)\leftarrow(t',\rvec x').
\end{equation}
This, however, is nothing but the \emph{Feynman interpretation of antimatter}. Thus, in the framework of QET, particles and antiparticles appear as different aspects of one and the same entity. Let us stick to the name ``particle'' for this entity that propagates from one event to another. If this propagation happens to be backwards in time, then the observer will \emph{interpret} it as the propagation of an antiparticle. In section~\ref{SeqFreeCase}, we will identify the detection of antiparticles with negative-energy events.

\subsection{Gauge transformation}

Following postulate~\ref{Evolution}, the physical behaviour of the particle is completely defined by the kinetic 4-momentum $\hat{\mvec\wp}$ whose components obey the commutation relations
\begin{eqnarray}
	[\hat \wp^\alpha,\hat \wp^\beta]&=&-i\hat F^{\alpha\beta}\label{PP}\\{}
	[\hat x^\alpha,\hat \wp^\beta]&=&i g^{\alpha\beta}\label{xP}.
\end{eqnarray}
Therefore, any transformation
\begin{eqnarray}
	\hat{\mvec\wp}\mapsto\hat{\mvec\wp'},
\end{eqnarray}
that leaves the commutation relations~(\ref{PP},\ref{xP}) invariant, does not affect the physical behaviour of the system. Let us call such a transformation a \emph{gauge transformation}.
Defining 
\begin{equation}
	\hat{\mvec\eta}:=\hat{\mvec\wp'}-\hat{\mvec\wp},
\end{equation}
we find
\begin{eqnarray}
	[\hat\eta^\alpha,\hat x^\beta]&=&[\hat \wp'^\alpha,\hat x^\beta]
		-[\hat \wp^\alpha,\hat x^\beta]\\
		&=&i g^{\alpha\beta}-i g^{\alpha\beta}=0,
\end{eqnarray}
and thus $\hat{\mvec\eta}$ must be diagonal in $\mvec x$,
\begin{equation}
	\hat{\mvec\eta}=\mvec\eta(\hat{\mvec x})
	=\int d^4x\,\mvec\eta(\mvec x)\,|\mvec x\rangle\langle\mvec x|.
\end{equation}
We then find
\begin{eqnarray}
	[\hat \wp^\alpha,\hat\eta^\beta]
		&=&[\hat p^\alpha,\hat\eta^\beta]-[\hat A^\alpha,\hat\eta^\beta]\\
		&=&[\hat p^\alpha,\hat\eta^\beta]\\
		&=&i\partial^\alpha\eta^\beta(\hat{\mvec x}).\label{etaP}
\end{eqnarray}
We further see that
\begin{eqnarray}
	[\hat P'^\alpha,\hat P'^\beta]&=&[\hat \wp^\alpha,\hat \wp^\beta]
		+[\hat\eta^\alpha,\hat \wp^\beta]+[\hat \wp^\alpha,\hat\eta^\beta]
		\\
		&\stackrel!=&[\hat \wp^\alpha,\hat \wp^\beta],
\end{eqnarray}
and thus
\begin{eqnarray}
	[\hat \wp^\alpha,\hat\eta^\beta]&=&[\hat \wp^\beta,\hat\eta^\alpha]
\end{eqnarray}
or equivalently, by using~\eqref{etaP},
\begin{equation}
	i\partial^\alpha\eta^\beta(\hat{\mvec x})
		=i\partial^\beta\eta^\alpha(\hat{\mvec x}).
\end{equation}
This implies that $\hat{\mvec\eta}$ is of the form
\begin{equation}
	\hat{\mvec\eta}=\mvec\partial\chi(\hat{\mvec x})
		=\int d^4x\,\mvec\partial\chi(\mvec x)
		|\mvec x\rangle\langle\mvec x|\label{etachi}
\end{equation}
for some differentiable real-valued function $\chi$ over $\mathbbm R^4$. A gauge transformation is therefore of the form
\begin{equation}\label{Pgauge}
	\hat{\mvec\wp}\mapsto\hat{\mvec\wp}+\hat{\mvec\eta},
\end{equation}
with $\hat{\mvec\eta}$ given by~\eqref{etachi}.
Because of the definition~\eqref{A}, the transformation~\eqref{Pgauge} can alternatively be conceived as a gauge transformation of the vector potential $\hat{\mvec A}$ according to
\begin{equation}\label{Agauge}
	\hat{\mvec A}\mapsto\hat{\mvec A}{}'=\hat{\mvec A}-\hat{\mvec\eta},
\end{equation}
in which case the operator $\hat{\mvec\eta}$ would correspond to a \emph{freedom of gauge} of the 4-potential $\hat{\mvec A}$. 
The kinetic 4-momentum $\hat{\mvec\wp}$ generates a modified translation,
\begin{eqnarray}
	e^{i\,d\mvec a\cdot\hat{\mvec\wp}}|\mvec x\rangle
		&=&|\mvec x+d\mvec a\rangle+i\,d\mvec a
		\cdot\mvec A(\mvec x)|\mvec x\rangle,
\end{eqnarray}
so that the transformed event wave function $\psi(\mvec x)=\langle\mvec x|\psi\rangle$ reads
\begin{eqnarray}
	e^{i\,d\mvec a\cdot\hat{\mvec\wp}}\psi(\mvec x)
		&=&\psi(\mvec x-d\mvec a)-i\,d\mvec a
		\cdot\mvec A(\mvec x)\psi(\mvec x).
\end{eqnarray}
One sees that besides translating the event wave function to another spacetime point, the above transformation also modifies the the amplitude by a certain amount.
In spacetime representation the kinetic 4-momentum becomes
\begin{equation}
	\hat{\mvec\wp}=i\mvec D,
\end{equation}
where 
\begin{equation}
	\mvec D:=\mvec\partial+i\mvec A(\mvec x)
\end{equation}
is the so-called \emph{covariant derivative}, because it transforms covariantly under the \emph{local gauge transformation}
\begin{eqnarray}
	\mvec A(\mvec x)&\mapsto&\mvec A'(\mvec x)
		=\mvec A(\mvec x)-\mvec\partial\chi(\mvec x)\\
	\psi(\mvec x)&\mapsto&\psi'(\mvec x)=e^{i\chi(\mvec x)}\psi(\mvec x),
\end{eqnarray}
which means that
\begin{eqnarray}
	D\psi(\mvec x)&\mapsto&D'\psi'(\mvec x)
		=e^{i\chi(\mvec x)}D\psi(\mvec x).
\end{eqnarray}
Thus, in the framework of QET, gauge transformations of the vector potential and the wave function are equivalent to gauge transformations of the observable $\hat{\mvec\wp}$.

In the free case we have $\hat F^{\alpha\beta}=0$ and therefore the kinetic 4-momentum $\hat{\mvec\wp}$ obeys the same commutation relations as the total 4-momentum $\hat{\mvec p}$. Thus there is a gauge transformation that maps $\hat{\mvec\wp}\mapsto\hat{\mvec p}$, so that in the free case the kinetic 4-momentum $\hat{\mvec\wp}$ can be physically identified with the total 4-momentum $\hat{\mvec p}$.

\section{The free case}\label{SeqFreeCase}

Let us now consider the free case, that is, the case where $\hat F^{\alpha\beta}=0$. The kinetic 4-momentum can in this case be gauge transformed into the total 4-momentum, $\hat{\mvec \wp}\mapsto\hat{\mvec p}$, and so the propagator becomes $\hat{\cal G}\mapsto\hat G$, where
\begin{equation}
	\hat G=\delta(\hat{\mvec p}{}^2-m^2)\label{freeG}
\end{equation} 
is called the \emph{free propagator}.
The natural unit of the mass $m$ is therefore given by
\begin{equation}
	[m]=[\sqrt{\mvec p^2}]=\frac1{\rm m},
\end{equation}
where we recall that ${\rm m}$ means ``meter''.
The family of spinless particles can so far be classified as follows:
\begin{eqnarray}
	m^2>0 &:& \text{massive particles}\\
	m=0 &:& \text{massless particles}\\
	m^2<0 &:& \text{imaginary-mass particles (tachyons)}.
\end{eqnarray}
The case $m^2>0$ actually divides into the cases $m>0$ and $m<0$. Both type of particles are governed by the same propagator, so they show the same behaviour. However, it seems that such negative-mass particles are not physically realized in the accessible part of the universe, so we will exclude this case in the following.
Similarily, the tachyonic case $m^2<0$ is a solution which does not seem to correspond to a physically realized type of particle. 
The $m=0$ case deserves a special treatment which we will also not provide here. 
Altogether, let us in the following restrict to the case of positive-mass particles, $m>0$.

According to postulate~\ref{Evolution}, a transition $\psi\rightarrow\phi$ is only allowed if the self-amplitudes of $\psi$ and $\phi$ do not vanish, 
\begin{equation}
	\tau(\phi,\phi)\neq0,\quad\tau(\psi,\psi)\neq0,
\end{equation}
that is, $\psi$ and $\phi$ must be \emph{physically allowed} quantum events.
For the free case an event $\psi$ is physically allowed if
\begin{equation}
	\langle\psi|\delta(\hat{\mvec p}{}^2-m^2)|\psi\rangle\neq0.
\end{equation}
The free propagator can be transformed as
\begin{eqnarray}
	\hat G &=&
		\delta(\hat{\mvec p}^2-m^2)
		=\int d^4p\,\delta(\mvec p^2-m^2)|\mvec p\rangle\langle\mvec p|\\
		&=&\int dE\int d^3p\,\frac1{2E_p}\big\{\delta(E-E_p)
			+\delta(E+E_p)\big\}|E,\rvec p\rangle\langle E,\rvec p|\\
		&=&\int d^3p\,\frac1{2E_p}
			|E_p,\rvec p\rangle\langle E_p,\rvec p|
		+ \int d^3p\,\frac1{2E_p}
			|\!-\!E_p,\rvec p\rangle\langle -\!E_p,\rvec p|,\label{GE}
\end{eqnarray}
where
\begin{equation}
	E_p :=\sqrt{\rvec p^2+m^2}.
\end{equation}
Thus any physically allowed event $\psi$ must have support on the upper or lower mass shell.
Therefore it is reasonable to categorize free particles according to the sign of their energy. 
Introduce the two projectors onto the positive- and negative energy subspace,
\begin{equation}
	\hat\Pi_\pm:=\theta(\pm\hat E),
\end{equation}
where 
\begin{equation}
	\theta(x):=\begin{cases}1&; x>0\\0&;x<0\end{cases}
\end{equation}
is the Heaviside step function, then
\begin{equation}
	\hat\Pi_+\hat\Pi_-=0,\quad\hat\Pi_++\hat\Pi_-=\mathbbm 1.
\end{equation}
Now consider a physically allowed event $\psi_+\in{\cal E}$ with only positive-energy contribution, 
\begin{equation}
	\langle\psi_+|\hat\Pi_-|\psi_+\rangle=0.
\end{equation}
The orbit of such an event reads
\begin{eqnarray}
	\Psi_+(t,\rvec x)
		&=&\langle t,\rvec x|\hat G|\psi_+\rangle\\
		&=&\langle t,\rvec x|\int d^3p\,\frac1{2E_p}
			|E_p,\rvec p\rangle\langle E_p,\rvec p|\psi_+\rangle\\
		&=&\int d^3p\,\frac1{2E_p} e^{-i(E_p t-\rvec p\rvec x)}\psi_+(E_p,\rvec p),
\end{eqnarray}
which equals a relativistic positive-energy wave package propagating forwards in time and space. Next, consider a physically allowed event $\psi_-\in{\cal E}$ with only negative-energy contribution, 
\begin{equation}
	\langle\psi_-|\hat\Pi_+|\psi_-\rangle=0,
\end{equation}
whose orbit reads
\begin{eqnarray}
	\Psi_-(t,\rvec x)
		&=&\langle t,\rvec x|\hat G|\psi_-\rangle\\
		&=&\langle t,\rvec x|\int d^3p\,\frac1{2E_p}
			|\!-\!E_p,\rvec p\rangle\langle\!-\!E_p,\rvec p|\psi_-\rangle\\
		&=&\int d^3p\,\frac1{2E_p} e^{+i(E_p t-\rvec p\rvec x)}\psi_-(-E_p,-\rvec p),
\end{eqnarray}
which equals a negative-energy wave package propagating backwards in time and space. According to theorem~\ref{TheoCharge}, this cannot be distinguished from an antiparticle wave propagating \emph{forwards} in time and space.
Consequently, the negative energy event $\psi_-$ appears to the observer as the evidence of an antiparticle.

\begin{figure}[t]
	\[\includegraphics[width=0.9\textwidth]{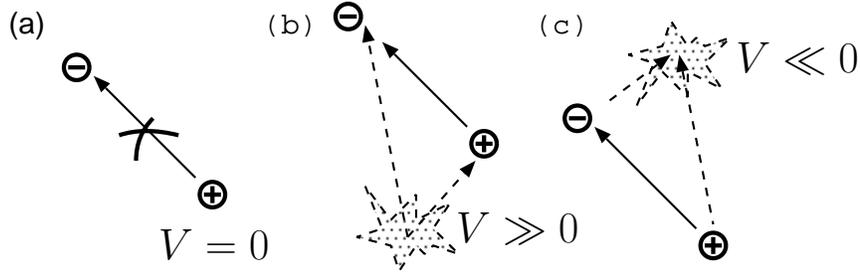}\]
		\vspace*{-0.5cm}
		\caption{\small (a): In the free case there is no transition from a positive-energy event (evidence of a particle) to a negative-energy event (evidence of an antiparticle). (b) and (c): If there is an adequate interaction potential, then such transition becomes possible, which the observer interprets as the creation or annihilation of a particle-antiparticle pair (dotted lines).}
		\label{pair}
\end{figure}

There can be no conversion of a free particle into its antiparticle, because
\begin{equation}
	\hat\Pi_-\hat G\hat\Pi_+ =0,
\end{equation}
as one can directly see by~\eqref{GE}.
Thus, an event $|\psi\rangle$ with only positive energy cannot cause an event $|\phi\rangle$ with only negative energy and vice versa.
This no longer holds true for the non-free case. Here it is possible that a positive-energy event causes a negative-energy event due to the interaction. Such transition appears to the observer as particle-antiparticle pair creation or annihilation (see Fig.~\ref{pair}), depending on the nature of the interaction potential.

\section{SI units and natural constants}

So far, the only unit that we need is the unit of \emph{distance}, arbitrarily defined as 1 \emph{meter}, and there are no natural contants. Let us now implement some additional units that are historically motivated. After this implementation the unit system will be called the \emph{SI unit system} and there will have appeared some constants which will be called \emph{natural constants}.

First, we introduce the \emph{second} ({\rm s}) as the unit of time. This is implemented by the replacement
\begin{equation}
	(t,\rvec x)\mapsto (ct,\rvec x),\label{ct}
\end{equation}
where the speed of light is set to
\begin{equation}
	c:=2.99792458\cdot 10^{8}{\rm m/s}.
\end{equation}
Consequently, 1 second is the time that light needs to travel a distance of 299,792,458 meters. The time $t$ is measured in seconds while the components of a spacetime point 
\begin{equation}
	\mvec x=(ct,\rvec x),
\end{equation}
are still measured in meters. In analogy to the temporal component of $\mvec x$ we redefine the temporal component of the total 4-momentum $\mvec p$ according to
\begin{equation}
	(E,\rvec p)\mapsto (E/c,\rvec p),
\end{equation}
so that $\mvec p\mvec x=Et-\rvec p\rvec x$ remains unchanged.
Next, we introduce the \emph{Coulomb} ({\rm C}) as the unit of charge. This is implemented by replacing the potential 4-momentum by
\begin{equation}
	\mvec A\mapsto \frac ec\mvec A,
\end{equation}
where the quantity
\begin{equation}
	e:=1.60217646\cdot 10^{-19}\, {\rm C}
\end{equation}
is called the \emph{elementary charge}. The kinetic 4-momentum thus obtains the form
\begin{equation}
	\mvec\wp=\mvec p-\frac ec\mvec A.
\end{equation}
Next, we introduce the \emph{Joule} ({\rm J}) as the unit of energy.
This is implemented by replacing the kinetic 4-momentum by
\begin{equation}
	\mvec\wp\mapsto\frac1\hbar\mvec\wp,
\end{equation}
where the quantity
\begin{equation}
	\hbar:=1.05457148\cdot 10^{-34}\,{\rm J\,s}
\end{equation}
is called the \emph{Planck constant}. Because of this replacement, the canonical commutation relations become
\begin{eqnarray}
	[\frac1\hbar\hat \wp^\alpha,\hat x^\beta]&=&ig^{\alpha\beta},
\end{eqnarray}
and thus
\begin{equation}
	[\hat \wp^\alpha,\hat x^\beta]=i\hbar g^{\alpha\beta}.
\end{equation}
Furthermore, the replacement gives rise to
\begin{equation}
	e^{-i\mvec p\mvec x}\mapsto e^{-\frac i\hbar\mvec p\mvec x}.
\end{equation}
In order for the 4-vector $\frac1\hbar\mvec p=\frac1\hbar(\mvec\wp-\frac ec\mvec A)$ to have the unit of ${\rm 1/m}$, so that $\frac1\hbar\mvec p\mvec x$ remains unit-free, the unit of the components of $\mvec p$ must be 
\begin{equation}
	[p^\alpha]=\frac{\rm J\,s}{\rm m}.
\end{equation}
Because $\frac ec\mvec A$ has the same units as $\mvec p$, we have
\begin{eqnarray}
	[A^\alpha]&=&\frac ce[p^\alpha]\\	
		&=&\frac{\rm m/s}{\rm C}\frac{\rm J\,s}{\rm m}\\
		&=&\frac{\rm J}{\rm C}, 
\end{eqnarray}
which coincides with the unit ``Volt''.
Up to here, the replacements induce the modification of the propagator according to
\begin{eqnarray}
	\delta(\hat{\mvec\wp}{}^2-m^2)
		&\mapsto&\delta\left(\left(\frac1\hbar\hat{\mvec\wp}\right)^2-m^2\right)\\
		&=&\hbar^2\,\delta(\hat{\mvec\wp}{}^2-\hbar^2 m^2),
\end{eqnarray}
where we have used that $\delta(cf(x))=\frac1c\delta(f(x))$ for any constant $c>0$.

Lastly, we introduce the \emph{kilogramm} ({\rm kg}) as the unit of mass. This is implemented by the replacement
\begin{equation}
	m\mapsto \frac{mc}{\hbar},
\end{equation}
so that the propagator becomes
\begin{equation}
	\hbar^2\delta(\hat{\mvec\wp}{}^2-\hbar^2 m^2)
		\mapsto\hbar^2\delta(\hat{\mvec\wp}{}^2-m^2c^2).
\end{equation}
We therefore find that the unit of mass, the kilogramm, obeys
\begin{eqnarray}
	[m]&\equiv&{\rm kg}\\
		&=&\left[\sqrt{\frac{\mvec p^2}{c^2}}\right]
		=\sqrt{\frac{\rm(J\,s/m)^2}{\rm(m/s)^2}}\\
		&=&\frac{\rm J\,s^2}{\rm m^2}.
\end{eqnarray}
From where we obtain 
\begin{equation}
	{\rm J}=\frac{\rm kg\,m^2}{\rm s^2}.
\end{equation}
Thus in terms of $({\rm kg, m, s, C})$, the units of $\mvec x$, $\mvec p$ and $\mvec A$ are given by
\begin{eqnarray}
	[x^\alpha]&=&{\rm m}\\~
	[p^\alpha]&=&\frac{\rm kg\,m}{\rm s}\\~
	[A^\alpha]&=&\frac{\rm kg\,m^2}{\rm s^2\,C}.
\end{eqnarray}
Altogether, the conversion from natural units to the SI unit system is performed by the replacements
\begin{eqnarray}
	(t,\rvec x)&\mapsto&(ct,\rvec x)\\
	(E,\rvec p)&\mapsto&(E/c,\rvec p)\\
	\mvec A&\mapsto&\frac ec\mvec A\\
	\mvec\wp&\mapsto&\frac1\hbar\mvec\wp\\
	m&\mapsto&\frac{mc}{\hbar}.
\end{eqnarray}
In SI units, the 4-position and the kinetic 4-momentum respectively read
\begin{equation}
	\mvec x=\begin{pmatrix}ct\\ \rvec x\end{pmatrix},\qquad
	\mvec\wp=\begin{pmatrix}E/c-\frac ec {\sf U}\\\rvec p-\frac ec\rvec A\end{pmatrix},
\end{equation}
the position-momentum brackets are given by
\begin{equation}
	\langle\mvec x|\mvec p\rangle=\frac1{\sqrt{2\pi\hbar}^4}e^{-\frac i\hbar\mvec p\mvec x},
\end{equation}
and the propagator becomes
\begin{eqnarray}
	\hat{\cal G}=\hbar^2\delta(\hat{\mvec\wp}{}^2-m^2c^2).
\end{eqnarray}

\section{The limit to standard Quantum Mechanics}

\subsection{Non-relativistic limit}

Now we will  perform the non-relativistic limit and find that in this limit the orbit wave function $\Psi(t,\rvec x)=\langle t,\rvec x|\hat{\cal G}|\psi\rangle$ obeys the familiar time-dependent Schr\"odinger equation. 
In order to perform the non-relativistic limit we switch to the SI unit system and rescale the kinetic energy so that the positive rest energy is set to zero. After that, we neglect terms of order $1/c^2$ and higher. This leads us us to the nonrelativistic particle approximation. The rescaling of the kinetic energy is done by remapping
\begin{equation}
	\wp^0\mapsto \wp^0+mc.
\end{equation}
This transforms the kinetic 4-momentum into
\begin{eqnarray}
	\hat{\mvec\wp}		
		&=&\begin{pmatrix}\frac1c(\hat {E}-e\hat  {\sf U})+mc\\
		\hat{\rvec p}-\frac ec\hat{\rvec A}\end{pmatrix}.
\end{eqnarray}
From there we obtain
\begin{eqnarray}
	\hat{\mvec\wp}{}^2&=&\left(\frac1c(\hat {E}-e\hat  {\sf U})+mc\right)^2
		-\left(\hat{\rvec p}-\frac ec\hat{\rvec A}\right)^2\\
		&=&\frac1{c^2}(\hat {E}-e\hat  {\sf U})^2+2m(\hat {E}-e\hat  {\sf U})
		+m^2c^2
		-\left(\hat{\rvec p}-\frac ec\hat{\rvec A}\right)^2.
\end{eqnarray}
Next, we neglect terms that are of order $1/c^2$ or higher.
The above expression then simplifies to
\begin{eqnarray}
	\hat{\mvec\wp}{}^2&\rightarrow& 2m(\hat {E}-e\hat  {\sf U})+m^2c^2
		-\left(\hat{\rvec p}-\frac ec\hat{\rvec A}\right)^2.
\end{eqnarray}
Therefore, the propagator is in the nonrelativistic limit given by
\begin{eqnarray}
	\hat{\cal G}&=&\hbar^2\delta(\hat{\mvec\wp}{}^2-m^2c^2)\\
		&\rightarrow&\hbar^2\delta\left(2m(\hat {E}-e\hat  {\sf U})
		-\left(\hat{\rvec p}
		-\frac ec\hat{\rvec A}\right){}^2\right)\\
		&=&\frac{\hbar^2}{2m}\delta\left(\hat E-e\hat {\sf U}-\frac1{2m}\left(\hat{\rvec p}
		-\frac ec\hat{\rvec A}\right){}^2\right),
\end{eqnarray}
which can be brought into the form
\begin{equation}\label{Gnonrel}
	\hat{\cal G}=\frac{\hbar^2}{2m}\delta(\hat{E}-\hat H),
\end{equation}
where the operator
\begin{equation}\label{H}
	\hat H:=\frac1{2m}\left(\hat{\rvec p}
		-\frac ec\hat{\rvec A}\right)^2+e\hat  {\sf U}
\end{equation}
is identified as the \emph{Hamiltonian} of the system. It can be brough into the form
\begin{equation}\label{Hdt}
	\hat H=\int dt\,|t\rangle\langle t|\otimes\hat H(t),
\end{equation}
where
\begin{equation}\label{Ht}
	\hat H(t) = \frac1{2m}\left(\hat{\rvec p}
		-\frac ec\hat{\rvec A}(t)\right)^2+e\hat  {\sf U}(t)
\end{equation}
acts on the factor space ${\cal H}=L^2(\mathbb R^3)$ of standard quantum mechanics.
The nonrelativistic propagator~\eqref{Gnonrel} obeys the equation
\begin{equation}
	(\hat E-\hat H)\hat{\cal G}=0,
\end{equation}
so that for some initial event $|\psi\rangle\in{\cal E}$ we obtain
\begin{eqnarray}
	\langle t,\rvec x|(\hat{E}-\hat H)\hat{\cal G}|\psi\rangle
		&=&\left(i\hbar\frac\partial{\partial t}-\hat H\right)
		\langle t,\rvec x|\hat{\cal G}|\psi\rangle=0,
\end{eqnarray}
where
\begin{equation}
	\hat H= \frac1{2m}\left(i\rvec\nabla+\frac ec\rvec A\right)^2
		+e {\sf U}
\end{equation}
is the time-dependent Hamiltonian of the system. 
Thus for the event orbit $\Psi(t,\rvec x)\equiv\langle t,\rvec x|\hat{\cal G}|\psi\rangle$ we obtain
\begin{equation}
	\left(i\hbar\frac\partial{\partial t}
		-\frac1{2m}\left(i\rvec\nabla+\frac ec\rvec A(t,\rvec x)\right)^2
		+e{\sf U}(t,\rvec x)\right)\Psi(t,\rvec x)
		=0,\label{Schroe}
\end{equation}
which coincides with the Schr\"odinger equation of a nonrelativistic particle exposed to an electromagnetic potential.
In terms of the propagator function $ {\cal G}(t,\rvec x;t',\rvec x')\equiv\langle t,\rvec x|\hat{\cal G}|t',\rvec x'\rangle$, we find
\begin{eqnarray}
	\langle t,\rvec x|(\hat E-\hat H)\hat{\cal G}|t',\rvec x'\rangle
		&=&\left(i\hbar\frac\partial{\partial t}-\hat H\right)
		 {\cal G}(t,\rvec x;t',\rvec x')
		=0,		
\end{eqnarray}
and thus ${\cal G}$ itself obeys the Schr\"odinger equation.
Let us temporarily assume that the Hamiltonian is constant in time.
In this case $\hat E$ and $\hat H$ commute and the propagator can be written as
\begin{equation}
	\hat{\cal G}=\frac{\hbar^2}{2m}\int dE\,\delta(E-\hat H)|E\rangle\langle E|,
\end{equation}
where $|E\rangle$ is an eigenvector of $\hat E\equiv\hat p^0$, so that
\begin{equation}
	|E,\rvec p\rangle=|E\rangle\otimes|\rvec p\rangle
\end{equation}
are the 4-position eigenvectors.
Now let us sandwich the propagator with time eigenvectors $|t\rangle$ appearing in the product form of the 4-position basis
\begin{equation}
	|t,\rvec x\rangle=|t\rangle\otimes|\rvec x\rangle.
\end{equation}
Then, using $\langle t|E\rangle=\frac1{\sqrt{2\pi\hbar}}e^{-\frac i\hbar Et}$, we obtain
\begin{eqnarray}
	\langle t|\hat{\cal G}|t'\rangle
		&=&\frac{\hbar}{4\pi m}\int dE\,e^{-\frac i\hbar E(t-t')}\delta(E-\hat H)\\
		&=&\frac{\hbar}{4\pi m}\,e^{-\frac i\hbar\hat H(t-t')}\\
		&=&\frac{\hbar}{4\pi m}\,\hat U(t-t'),
\end{eqnarray}
where $\hat U(t)=e^{-\frac i\hbar \hat Ht}$ is the unitary time evolution operator acting on the Hilbert space ${\cal H}=L^2(\mathbbm R^3)$ spanned by the position basis vectors $|\rvec x\rangle$. In case of a time-dependent Hamiltonian, the unitary time evolution operator is given by
\begin{equation}
	\hat U(t,t')={\cal T}e^{-\frac i\hbar\int_{t'}^{t}ds\,\hat H(s)},
\end{equation}
where ${\cal T}$ is the Dyson time ordering operator. So we can generally write the non-relativistic propagator in the form
\begin{equation}\label{GnonrelU}
	\hat{\cal G}=\kappa\int dt\int dt'\,|t\rangle\langle t'|\otimes\hat U(t,t').
\end{equation}
with $\kappa=\frac{\hbar}{4\pi m}$, and where $\hat U(t,t')$
obeys the Schr\"odinger equation
\begin{equation}
	\left(i\hbar\frac\partial{\partial t}-\hat H(t)\right)\hat U(t,t')=0.
\end{equation}
The transition amplitude between two events $|\psi\rangle$ and $|\phi\rangle$ thus becomes
\begin{eqnarray}
	\tau(\phi,\psi)&=&\langle\phi|\hat{\cal G}|\psi\rangle\\
		&=&\kappa\int dt\int dt'\,\langle\phi
		|\big(|t\rangle\langle t'|\otimes\hat U(t,t')\big)|\psi\rangle\\
		&=&\kappa\int dt\int dt'\,\langle\Phi_t|\hat U(t,t')|\Psi_{t'}\rangle,
\end{eqnarray}
where
\begin{eqnarray}
	|\Psi_t\rangle&\equiv&\langle t|\psi\rangle,\qquad
	|\Phi_t\rangle\equiv\langle t|\phi\rangle\label{timeproj}
\end{eqnarray}
are ``time-projected'' vectors in the factor space ${\cal H}=L^2(\mathbbm R^3)$. It should be noted that the norm of these vectors is a nontrivial function of $t$ and obeys
\begin{equation}
	\int dt\,\langle\Psi_t|\Psi_t\rangle
		=\int dt\,\langle\psi|t\rangle\langle t|\psi\rangle
		=\langle\psi|\psi\rangle<\infty.
\end{equation}
The family of vectors $|\Psi_t\rangle$ can therefore not be interpreted as the state trajectory $|\Psi(t)\rangle$ of the particle. This would not be meaningful anyway, because the dynamics of the particle is determined by the orbit $|\Psi\rangle=\hat{\cal G}|\psi\rangle$ and not by the event vector $|\psi\rangle$.

\subsection{Time-energy uncertainty}

Because the derived Hamiltonian~\eqref{Hdt} is diagonal in time, it commutes with the fundamental time operator,
\begin{equation}
	[\hat H,\hat t]=0.
\end{equation}
This goes in contrast to the time operators considered in other approaches. Also, it completely circumvents Pauli's counterargument against a time observable~\cite{Pauli26}. As a consequence, there is no time-energy uncertainty as long as energy is directly identified with the Hamiltonian $\hat H$.
On the other hand, we have a time-energy uncertainty relation between $\hat E$ and $\hat t$, because by construction
\begin{equation}
	[\hat E,\hat t]=i,
\end{equation}
and thus
\begin{equation}\label{DEDt}
	\Delta E\Delta t\geq\frac{1}{2},
\end{equation}
where $E$ indicates the eigenvalue of the fundamental energy operator $\hat E$.
How can this be understood? 

The time-energy uncertainty relation has always been a controversal issue~\cite{Aharonov61,Busch02,Brunetti02b,Eberly73}. 
There are theoretical considerations and experimental facts that strengthen an uncertainty relation between the measured energy of the system and the duration of this measurement. Such a time-energy uncertainty relation can also be derived in a quite general way when considering an interacting environmental system~\cite{Briggs00,Briggs01}.
On the other hand it has been shown by Aharonov and Bohm~\cite{Aharonov61} that there are particular cases where the duration of an energy measurement does not affect the uncertainty of the measurement result. This paradoxical situation can be solved~\cite{Aharonov01} if one distinguishes between the measurement of an apriori \emph{known} and an apriori \emph{unknown} Hamiltonian: 
In case the Hamiltonian is unknown it must be estimated by observing the dynamics of the system. The uncertainty of such estimation and the duration of the measurement obey a relation of the form~(\ref{DEDt}). On the other hand, if the Hamiltonian is explicitely known, then it is possible to design a measurement which circumvents a time-energy uncertainty relation, as shown in~\cite{Aharonov61}. 

In QET, the operator $\hat E$ represents the energy, but it is not identical to the Hamiltonian. In the nonrelativistic limit, the Hamiltonian shows up as exactly that operator that coincides with the energy of the particle's orbit $|\Psi\rangle=\hat{\cal G}|\psi\rangle$, because the orbit fullfills
\begin{equation}
	(\hat E-\hat H)|\Psi\rangle=0.
\end{equation}
The orbit is not a member of the Hilbert space ${\cal E}$ and thus cannot be regarded as an event, but it is the closest thing next to the ``trajectory'' of a particle, because its spacetime representation $\Psi(t,\rvec x)$ obeys the Schr\"odinger equation. Altogether, the energy is, like the momentum, a fundamental observable which is not dependent on external settings, and which fulfills an uncertainty relation with time, while the Hamiltonian is an operator that is dependent on external settings (the potentials), and that appears in the nonrelativstic limit as an operator whose eigenvalues coincide with the energy eigenvalues along the trajectory of the system.

\subsection{Sharp-time limit}

Let us in the non-relativistic approximation consider the transition amplitudes between two events $\psi$ and $\phi$ under the assumption that the time uncertainty of both events is neglegible, an approximation which we will call the \emph{sharp-time limit}.
We will find that in this limit the transition probabilities between two events take the same form as those of standard quantum mechanics.

The sharp-time limit is realized by considering the two involved events as sharply peaked at two distinct times $t_0$ and $t_1$.
The abstract representations of these two events are given by
\begin{eqnarray}
	|\psi\rangle&=&|t_0,\Psi\rangle\equiv|t_0\rangle\otimes|\Psi\rangle\\
	|\phi\rangle&=&|t_1,\Phi\rangle\equiv|t_1\rangle\otimes|\Phi\rangle,
\end{eqnarray}
Where $|\Psi\rangle,|\Phi\rangle$ are elements of the factor space ${\cal H}=L^2(\mathbbm R^3)$ that contains the ordinary quantum ``states''.
The time-projected vectors~\eqref{timeproj} become
\begin{eqnarray}
	|\Psi_t\rangle&=&\delta(t-t_0)|\Psi\rangle,\qquad
	|\Phi_t\rangle=\delta(t-t_1)|\Phi\rangle.
\end{eqnarray}
Then, using the non-relativistic propagator~\eqref{GnonrelU}, the amplitude for the transition $\psi\rightarrow\phi$ becomes
\begin{eqnarray}
	\tau(\phi,\psi)&=&\langle\phi|\hat{\cal G}|\psi\rangle\\
		&=&\langle t_1,\Phi|\hat{\cal G}|t_0,\Psi\rangle\\
		&=&\kappa\int dt\int dt'\langle t_1|t\rangle\langle t'|t_0\rangle
		\langle\Phi|\hat U(t,t')|\Psi\rangle\\
		&=&\kappa\int dt\int dt'\delta(t-t_1)\delta(t'-t_0)
		\langle\Phi|\hat U(t,t')|\Psi\rangle\\
		&=&\kappa\langle\Phi|\hat U(t_1,t_0)|\Psi\rangle.
\end{eqnarray}
Because $\hat U(t,t)=\mathbbm1$ for all $t\in\mathbbm R$, the transition probability for the process reads
\begin{eqnarray}
	P(\phi,\psi)&=&\frac{|\tau(\phi,\psi)|^2}
		{\tau(\phi,\phi)\tau(\psi,\psi)}\\
		&=&\frac{|\kappa|^2|\langle\Phi|\hat U(t_1,t_0)|\Psi\rangle|^2}
			{|\kappa|^2\langle\Phi|\hat U(t_1,t_1)|\Phi\rangle
			\langle\Psi|\hat U(t_0,t_0)|\Psi\rangle}\\
		&=&\frac{|\langle\Phi|\hat U(t_1,t_0)|\Psi\rangle|^2}
			{\langle\Phi|\Phi\rangle
			\langle\Psi|\Psi\rangle},
\end{eqnarray}
which coincides with the probability for the transition from the state $\Psi$ at time $t_0$ to the state $\Phi$ at time $t_1$, as it is obtained in standard nonrelativistic quantum mechanics.
It should be remarked that in standard quantum mechanics the time uncertainty is always assumed to be zero, because the system is necessarily in a certain state at a certain time, and there is no room for any ``time uncertainty''. In the framework of QET, a finite time uncertainty is obligatory because the vectors with zero time uncertainty, the time eigenvectors, are no members of the event Hilbert space ${\cal E}$ and thus do not represent physically possible events. Hence, it is no longer possible to say that a particle is in a certain state at a certain time.

Altogether, we have shown that in the nonrelativistic sharp-time limit, the formalism of QET yields the same predictions as those obtained from standard non-relativistic QM.

\section{Epilog}

The present paper is an outline of the basic concepts for a quantum mechanical theory that describes the kinematics of individual events rather than of individual states. Since this mechanical concept deviates from the standard concept, the theory is a mechanical theory alternative to the existing mechanical theories.

There are several interesting open tasks, some of which are to be presented in a subsequent paper:
\begin{enumerate}
\item
Investigate causality in the framework of QET.
\item
Elaborate a perturbation theory which allows the construction of Feynman-like diagrams.
\item
Include spin in the description.
\item
Work out a theory of multi-events, and investigate the connection to Quantum Field Theory.
\item
Think about how gravity could be integrated into QET.
\end{enumerate}
This work is the result of several years of reasoning and it has a moved history on its own.
I could hardly mention all arguments, counterarguments, associations, and controversies that have come along the way. 
I am thankful for fruitful discussions with other collegues, in particular with Jens Eisert, Timo Felbinger, and Juan Gonzalo Muga.
I am aware that the topic of the paper touches highly controversial issues, but there can be no scientific progress without sometimes taking a risk. I hope that the reader tries to understand the ideas presented in this paper, and does not retreat to a categorical denial because of the somewhat unorthodox presentation, or because of a different philosophical attitude. If the reader does not agree with certain conceptual and philosophical aspects, he or she may take the theory at least as a mathematical construct that provides a self-consistent description of a small but fundamental part of the physical world, based on some unusual assumptions about the nature of time and space.

\bibliography{QET3}
\bibliographystyle{unsrt}

\end{document}